\documentclass{aa} 

\usepackage{graphicx}
\usepackage{tabularx}
\usepackage{placeins}
\usepackage[rightcaption]{sidecap}
\usepackage[english]{babel}
\usepackage{array}
\usepackage{capt-of}
\usepackage[normalem]{ulem}
\usepackage{lscape}

\newcolumntype{L}[1]{>{\raggedright\let\newline\\\arraybackslash\hspace{0pt}}m{#1}}
\newcolumntype{C}[1]{>{\centering\let\newline\\\arraybackslash\hspace{0pt}}m{#1}}
\newcolumntype{R}[1]{>{\raggedleft\let\newline\\\arraybackslash\hspace{0pt}}m{#1}}

\usepackage{txfonts}
\usepackage[colorlinks=true, allcolors=blue]{hyperref}
\def\kms{km~s$^{-1}$}

\graphicspath{{folder/}{Figures/}}

\begin{document}

   \title{Methanol deuteration in high-mass protostars}


   \author{
          M. L. van Gelder\inst{1}
          \and
          J. Jaspers\inst{1}
          \and
          P. Nazari\inst{1}
          \and
          A. Ahmadi\inst{1}
          \and
          E. F. van Dishoeck\inst{1,2}
          \and 
          M. T. Beltr\'an\inst{3}
          \and \\
          G. A. Fuller\inst{4,5}
          \and
          \'A. S\'anchez-Monge\inst{5}
          \and
          P. Schilke\inst{5}
          }
\institute{
         Leiden Observatory, Leiden University, PO Box 9513, 2300RA Leiden, The Netherlands \\
         \email{vgelder@strw.leidenuniv.nl}
         \and
         Max Planck Institut f\"ur Extraterrestrische Physik (MPE), Giessenbachstrasse 1, 85748 Garching, Germany
         \and
         INAF-Osservatorio Astrofisico di Arcetri, Largo E. Fermi 5, 50125 Firenze, Italy
         \and
         Jodrell Bank Centre for Astrophysics, Department of Physics and Astronomy, University of Manchester, Oxford Road, Manchester, M13 9PL, UK
         \and
         I. Physikalisches Institut, Universit\"at zu K\"oln, Z\"ulpicher Str.77, 50937, K\"oln, Germany
          }

   \date{Received XXX; accepted XXX}

 
  \abstract
   {
   The deuteration of molecules forming in the ices such as methanol (CH$_3$OH) is sensitive to the physical conditions during their formation in dense cold clouds and can be  probed through observations of deuterated methanol in hot cores.
   }
   {
   The aim is to determine the D/H ratio of methanol for a large sample of 99 high-mass protostars and to link this to the physical conditions during the formation of methanol in the prestellar phases.
   }
   {
   Observations with the Atacama Large Millimeter/submillimeter Array (ALMA) containing transitions of CH$_3$OH, CH$_2$DOH, CHD$_2$OH, $^{13}$CH$_3$OH, and CH$_3^{18}$OH are investigated. 
   The column densities of CH$_2$DOH, CHD$_2$OH, and CH$_3$OH are determined for all sources, where the column density of CH$_3$OH is derived from optically thin $^{13}$C and $^{18}$O isotopologues. Consequently, the D/H ratio of methanol is derived taking statistical effects into account.
   }
   {
   Singly deuterated methanol (CH$_2$DOH) is detected at the $3\sigma$ level toward 25 of the 99 sources in our sample of the high-mass protostars. Including upper limits, the $\rm (D/H)_{CH_3OH}$ ratio inferred from $N_\mathrm{CH_2DOH}/N_\mathrm{CH_3OH}$ was derived for 38 of the 99 sources and varies between $\sim10^{-3}-10^{-2}$. 
   Including other high-mass hot cores from the literature, the mean methanol D/H ratio is $1.1\pm0.7\times10^{-3}$. 
   This is more than one order of magnitude lower than what is seen for low-mass protostellar systems ($2.2\pm1.2\times10^{-2}$).
	Doubly deuterated methanol (CHD$_2$OH) is detected at the $3\sigma$ level toward 11 of the 99 sources.  
	Including upper limits for 15 sources, the $\rm (D/H)_{CH_2DOH}$ ratios derived from $N_\mathrm{CHD_2OH}/N_\mathrm{CH_2DOH}$ are more than two orders of magnitude higher than $\rm (D/H)_{CH_3OH}$ with an average of $2.0\pm0.8\times10^{-1}$ which is similar to what is found for low-mass sources.
   Comparison with literature GRAINOBLE models suggests that the high-mass prestellar phases are either warm ($>20$~K) or live shorter than the free-fall timescale. In contrast, for low-mass protostars, both a low temperature of $<15$~K and a prestellar phase timescale longer than the free-fall timescale are necessary.
   }
   {
   The $\rm (D/H)_{CH_3OH}$ ratio drops by more than an order of magnitude between low-mass and high-mass protostars due to either a higher temperature during the prestellar phases or shorter prestellar phases. However, successive deuteration toward CHD$_2$OH seems equally effective between low-mass and high-mass systems.
   }


   \maketitle
%

\section{Introduction}
\label{sec:introduction}
Isotopologues have proven to be vital in our understanding of the star and planet formation process. They allow for studying the most abundant species for which the emission originating from the main isotopologue is optically thick. Moreover, the sensitivity of isotopologue ratios to the physical conditions such as temperature and ultraviolet (UV) radiation has proven key in understanding the molecular journey during the entire star formation process \citep[see e.g., reviews by][]{Caselli2012,Tielens2013,Ceccarelli2014}. One of the most studied isotopologues are those that contain deuterium (D). These deuterated molecules are suggested to form already in the cold prestellar phases \citep[e.g.,][]{vanDishoeck1995,Caselli2012,Ceccarelli2014}. Especially for molecules such as methanol (CH$_3$OH) that form on the surfaces of dust grains in dense cores, D/H fractionation ratios up to 10\% are found toward low-mass protostars \citep[e.g.,][]{Bianchi2017_SVS13A,Bianchi2017_HH212,Taquet2019,vanGelder2020}, more than four orders of magnitude larger than the canonical D/H ratio derived for the local interstellar medium (ISM) of of $\sim2\times10^{-5}$ \citep{Linsky2006,Prodanovic2010}. It is thus key to understand the deuterium fractionation process in the earliest phases of star formation.

The gaseous atomic D/H ratio can be increased in the prestellar phases through the exothermic reaction \citep{Watson1974,Aikawa1999,Ceccarelli2014},
\begin{align}
\rm H_3^+ + HD \rightleftharpoons H_2D^+ + H_2 + \Delta E,
\label{eq:H2D+_formation}
\end{align}
where $\rm \Delta E = 232$~K. Since in the cold ($\lesssim20$~K) prestellar cores the backward reaction in Eq.~\eqref{eq:H2D+_formation} is less efficient, H$_2$D$^+$ is enhanced and the atomic D/H ratio in the gas phase can be effectively increased through dissociative recombination of $\rm H_2D^+$ with free electrons. Moreover, gaseous CO is the main destructor of $\rm H_3^+$ and $\rm H_2D^+$ \citep{Brown1989,Roberts2003} and thus the heavy CO freeze-out in dense ($\gtrsim10^{4}$~cm$^{-3}$) prestellar cores additionally stimulates the increase of the gaseous atomic D/H ratio. In turn, the enhanced atomic D/H ratio in the gas can translate into a higher D/H ratio of molecules forming in the ices \citep{Tielens1983,Nagaoka2005}. Measuring the deuteration of molecules that form in the ices is thus a powerful tool to determine the physical conditions such as density (e.g., CO freeze-out) and temperature during their formation.

Methanol forms on the surfaces of dust grains in dense prestellar phases through the hydrogenation of CO ice \citep[e.g.,][]{Watanabe2002,Fuchs2009} and reactions between its grains-surface products \citep[e.g., H$_2$CO and CH$_3$O][]{Simons2020,Santos2022} and is therefore expected to exhibit a high D/H ratio. This is in strong contrast to, for example, water for which the bulk of the ice is formed in the warmer translucent cloud phase leading to a rather low overall HDO/H$_2$O ratio \citep[$\lesssim0.1$\%;][]{Persson2014,Furuya2016,Jensen2019,vantHoff2022}. The sensitivity of the methanol deuteration process to temperature was investigated by \citet{Bogelund2018} using the GRAINOBLE gas-grain chemical model \citep{Taquet2012,Taquet2013,Taquet2014}, finding a strong correlation between the D/H ratio of methanol and the formation temperature. Moreover, \citet{Taquet2019} showed that the timescale of the prestellar phase is highly relevant for methanol deuteration.

Methanol and its (deuterated) isotopologues are readily observed as they desorb from the dust grains. Mono deuterated methanol, CH$_2$DOH and CH$_3$OD, have been observed in the warm inner regions of both low-mass and high-mass protostellar systems \citep[e.g.,][]{Fuente2014,Belloche2016,Bogelund2018,vanGelder2020,vanderWalt2021}. Similarly, both doubly and triply deuterated methanol have been detected in hot cores \citep[e.g.,][]{Parise2002,Parise2004,Bianchi2017_SVS13A,Drozdovskaya2022,Ilyushin2022}. Moreover, CH$_2$DOH has also been detected in both low-mass prestellar cores \citep[e.g.,][]{Bizzocchi2014,Lattanzi2020,Ambrose2021} and high-mass starless cores \citep[e.g.,][]{Fontani2015}. Across this mass and evolutionary range, the D/H ratio of singly deuterated methanol varies orders of magnitude. The D/H ratio is on the order of 10\% for low-mass prestellar cores, low-mass protostars, and comets \citep[e.g.,][]{Bianchi2017_SVS13A,Bianchi2017_HH212,Jorgensen2018, Taquet2019,Manigand2020,vanGelder2020,Lattanzi2020,Ambrose2021,Drozdovskaya2021}. Interestingly, successive deuteration toward CHD$_2$OH and CD$_3$OH seems to be quite effective in low-mass protostars \citep[about 15--25\%;][]{Drozdovskaya2022,Ilyushin2022}. On the other hand, the D/H ratio derived from CH$_2$DOH is as low as $0.1-0.01$\% for high-mass starless cores and high-mass protostars \citep{Fontani2015,Neill2013,Belloche2016,Bogelund2018}. However, the sample of high-mass protostars for which reliable and interferometrically derived methanol D/H ratios are available (e.g., Orion~KL, Sgn~B2(N2), NGC~6334I) remains small compared to the low-mass sources ($\sim20$ sources). Furthermore, no interferometric detections of CHD$_2$OH in high-mass sources have been presented thus far.

In this work, the methanol D/H ratios are derived for an additional 99 high-mass sources based on ALMA observations of CH$_2$DOH, CHD$_2$OH, CH$_3$OH $^{13}$CH$_3$OH, CH$_3^{18}$OH. In Sect.~\ref{sec:methodology}, the observations and derivation of the column densities are explained. The resulting D/H ratios of CH$_3$OH and CH$_2$DOH are presented in Sect.~\ref{sec:results}. In Sect.~\ref{sec:discussion}, the methanol D/H ratios derived for our high-mass sources are compared to their low-mass counterparts and prestellar phases. Furthermore, through comparison with the GRAINOBLE models computed by \citet{Bogelund2018} and \citet{Taquet2019}, the effect of physical conditions on the methanol D/H ratio is discussed. Our main conclusions are listed in Sect.~\ref{sec:conclusion}.

\section{Methodology}
\label{sec:methodology}
\subsection{Observations}
\label{subsec:observations}
The dataset analyzed in this work was taken from the ALMA Evolutionary study of High Mass Protocluster Formation in the Galaxy (ALMAGAL) survey (2019.1.00195.L; PI: S. Molinari) that targeted over 1000 dense clumps with $M>500$~M$_\odot$ based on the {\it Herschel} Hi-Gal survey \citep{Molinari2010,Elia2017,Elia2021}. The ALMAGAL survey covers frequencies from $\sim217$~GHz to $\sim221$~GHz with multiple configurations of ALMA down to $\sim0.1''$ resolution at a spectral resolution of between $\sim0.2-0.7$~\kms. In this work a subsample of 40 high-mass cores is selected based on high bolometric luminosity ($L_\mathrm{bol} > 1000$~L$_\odot$) and the sources being rich in lines from complex organic molecules (COMs) such as CH$_3$OH and CH$_3$CN. 
Only archival data with a beam smaller than 2$^{\prime\prime}$ ($\sim1000-5000$~au) that were public before February 2021 are included. 
This selection introduces a bias in our sample to line-rich sources and means that not all high-mass cores in the ALMAGAL survey are covered.
In the higher resolution ALMA data, the 40 Hi-Gal high-mass cores are resolved into in total 99 sources based on the continuum emission (labeled A, B, C, etc., see Appendix~\ref{app:Observational_details}). These 99 sources are all studied in this work and are the same as those that were analyzed by \citet{vanGelder2022}.
The data are pipeline calibrated and imaged with the Common Astronomy Software Applications\footnote{\url{https://casa.nrao.edu/}} \citep[CASA;][]{McMullin2007} version 5.6.1.
The angular resolution of the data ranges from $0.5-1.25''$, corresponding to about $\sim2500-10000$~au at the range of distances covered \citep[$2-12$~kpc;][]{Mege2021}, and the data have a sensitivity of $\sim0.2$~K. The ALMAGAL data cover several transitions of CH$_3$OH, four transitions of $^{13}$CH$_3$OH, nine transitions of CH$_3^{18}$OH, 21 transitions of CH$_2$DOH, and 22 transitions of CHD$_2$OH (see Appendix~\ref{app:CH3OH_transitions}). Also nine transitions of CD$_3$OH are covered \citep{Ilyushin2022}, but these are not detected toward any of the sources. No transitions of CH$_3$OD are covered in the observed frequency range. Both CD$_3$OH and CH$_3$OD are therefore not analyzed further in this paper.

\begin{figure*}[t]
\centering
\includegraphics[width=0.33\linewidth]{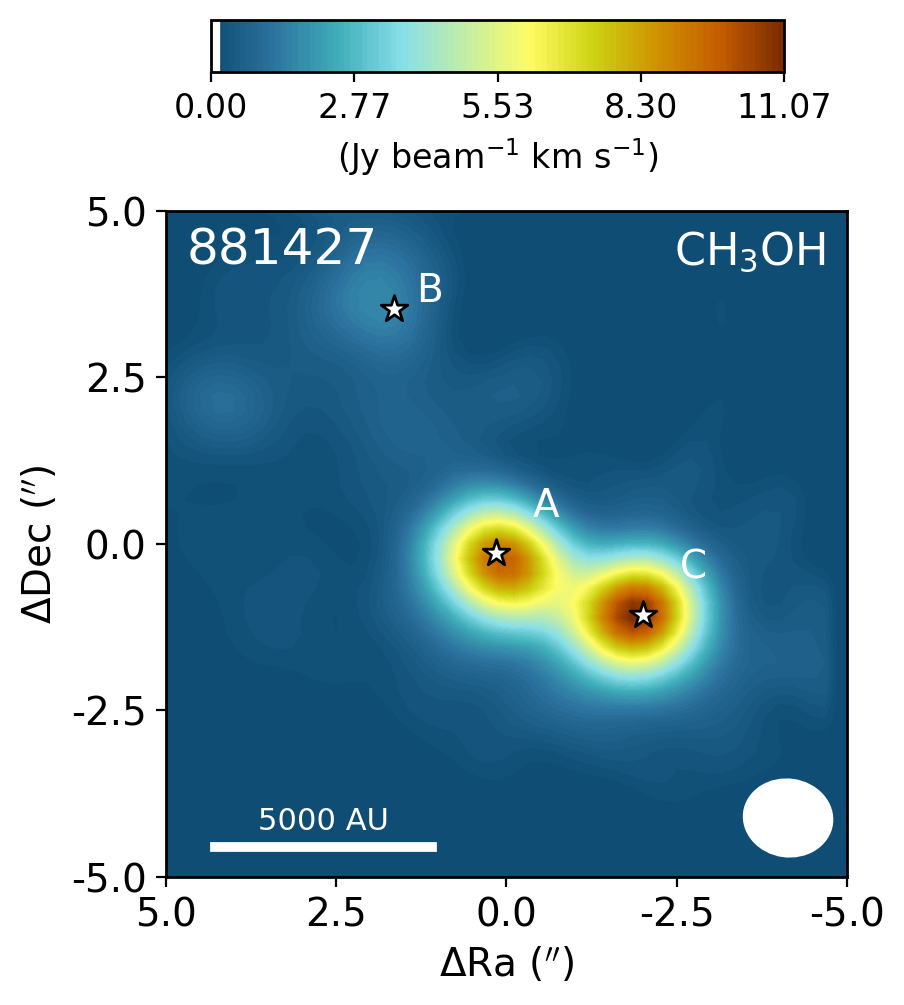}
\includegraphics[width=0.33\linewidth]{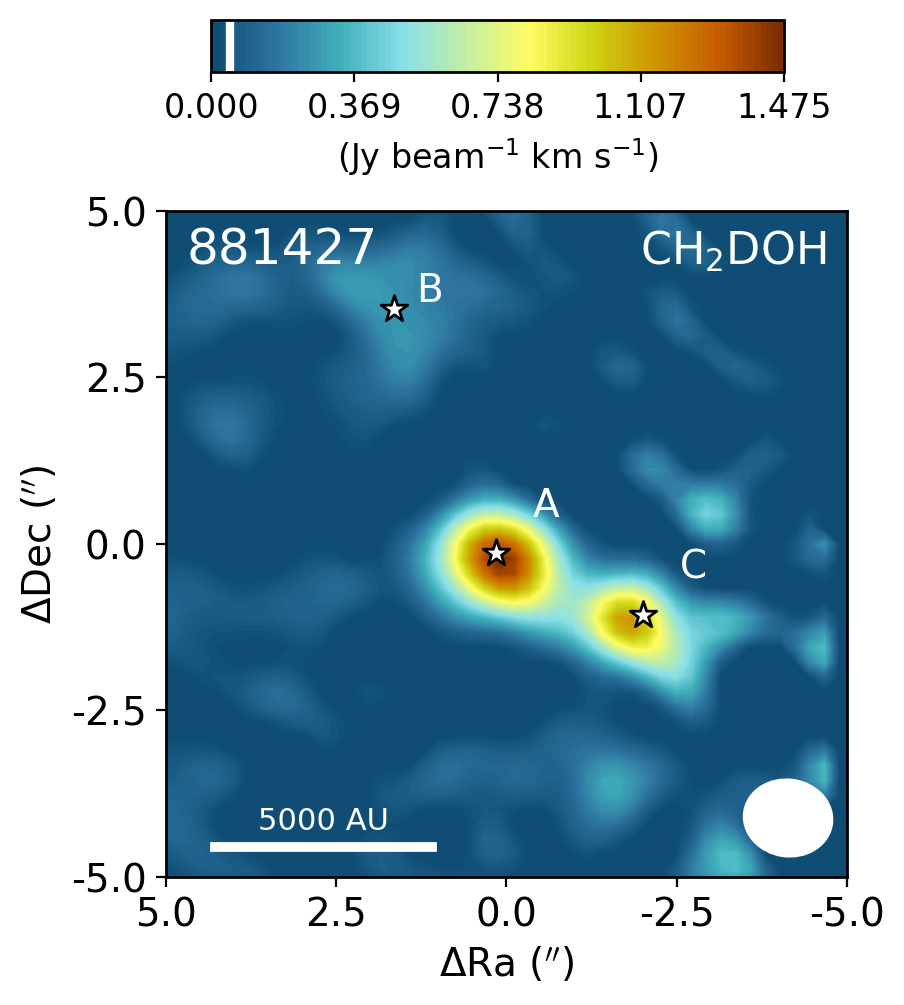}
\includegraphics[width=0.33\linewidth]{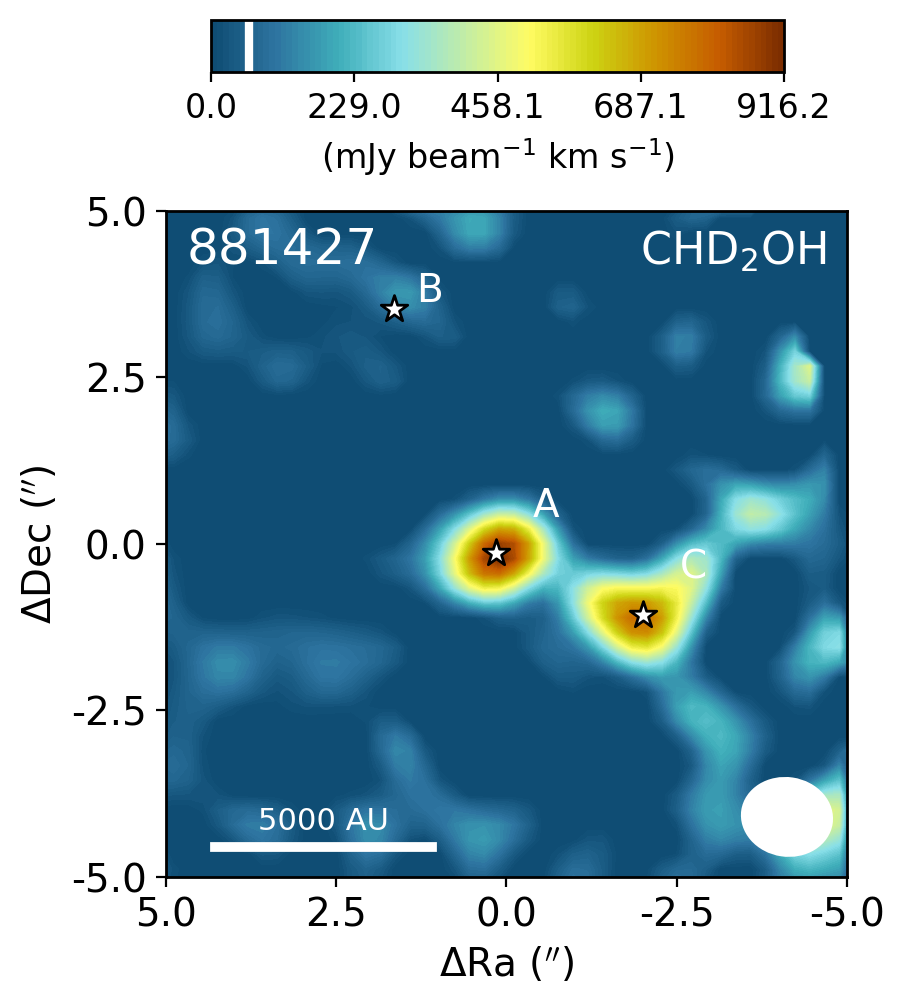}
\caption{Integrated intensity maps of the CH$_3$OH $8_{0,8} - 7_{1,6}$ ($E_\mathrm{up}=97$~K, left), CH$_2$DOH $17_{1,16}\,e_0-17_{0,17}\,e_0$ ($E_\mathrm{up} = 336$~K, middle), and CHD$_2$OH $7_{0,1}\,e_1-6_{1,1}\,e_1$ ($E_\mathrm{up} = 74$~K, right) lines for 881427. The color scale is shown on top of each image. The image is integrated over [-5,5]~\kms\ with respect to the $V_\mathrm{lsr}$ of source A. The white vertical line in the colorbar indicates the $3\sigma$ threshold. The source positions based on the continuum emission are indicated with the white stars. The white ellipse in the lower right of each image depicts the beam size and in the lower left a physical scale bar is displayed.}
\label{fig:mom0_881427}
\end{figure*}

Integrated intensity maps of the CH$_3$OH $8_{0,8} - 7_{1,6}$, CH$_2$DOH $17_{1,16}\,e_0-17_{0,17}\,e_0$, and CHD$_2$OH $7_{0,1}\,e_1-6_{1,1}\,e_1$ lines for the source 881427 are presented in Fig.~\ref{fig:mom0_881427}. The source 881427 hosts three nearby hot cores with varying line strengths and line widths and is a representative source of the rest of the sample. Whereas the emission of CH$_3$OH is often larger than the central beam, the emission of both CH$_2$DOH and CHD$_2$OH is generally confined within the central beam similar to the $^{13}$C and $^{18}$O isotopologues. The main exception for this is CH$_2$DOH $5_{1,5}\,e_0-4_{1,4}\,e_0$ ($E_\mathrm{up} = 36$~K) which often shows more extended emission. This is likely because the low upper energy level of this line is also sensitive to cold ($T\lesssim70$~K) material where methanol is nonthermally desorbed from the grains \citep[e.g.,][]{Perotti2020,Perotti2021}. To exclude the contribution of extended emission, this transition is not included in the analysis described below. 

The spectra are extracted from the peak pixel in the CH$_3$OH $8_{0,8} - 7_{1,6}$ ($E_\mathrm{up}=97$~K) integrated intensity maps for all sources that show this at the $>3\sigma$ level. This line is strongest transition of CH$_3$OH in the sample with $E_\mathrm{up}>70$~K (lines with lower $E_\mathrm{up}$ can suffer from contamination by the outflow or extended emission). For sources that do not show emission from the CH$_3$OH $8_{0,8} - 7_{1,6}$ line, spectra are extracted from the peak continuum pixel and only upper limits on the column densities of CH$_3$OH (and isotopologues) are derived. In the G323.7399-00.2617B cluster, which contains seven nearby cores, all spectra are extracted from the same positions as \citet{vanGelder2022}. 
In G023.3891+00.1851, the emission of CH$_2$DOH peaks offset by about half the beam ($\sim0.6''$) and therefore the spectrum is extracted from the peak of CH$_2$DOH $17_{1,16}\,e_0-17_{0,17}\,e_0$ ($E_\mathrm{up} = 336$~K). For all other sources, the the peak in CH$_3$OH coincides with the peaks of CH$_2$DOH and CHD$_2$OH.
For sources also included by \citet{Nazari2022_ALMAGAL}, our spectral extraction locations are the same as theirs, which are extracted from the peak position of the CH$_3$CN $12_{4} - 11_{4}$ integrated intensity maps, except for 721992 and G023.3891+00.1851 where the CH$_3$OH $8_{0,8} - 7_{1,6}$ and CH$_2$DOH $17_{1,16}\,e_0-17_{0,17}\,e_0$ emission peaks offset from the CH$_3$CN $12_{4} - 11_{4}$ emission by about $1''$. 
It is important to note that these spectral extraction positions are different by up to $1''$ from \citet{vanGelder2022} who extracted their spectra from the peak continuum pixel for all sources. Therefore, the column densities derived in this work may deviate from theirs. The reason why our spectra are extracted from the peak pixel of CH$_3$OH is to have the highest signal-to-noise in methanol lines and its isotopologues the extracted spectra. 

\subsection{Deriving the column densities}
The column densities of all methanol isotopologues are derived using the spectral analysis tool CASSIS\footnote{\url{http://cassis.irap.omp.eu/}} \citep{Vastel2015} under the assumption of local thermodynamic equilibrium (LTE). The line lists of CH$_3$OH, $^{13}$CH$_3$OH, and CH$_3^{18}$OH are taken from the CDMS catalog\footnote{\url{https://cdms.astro.uni-koeln.de/}} \citep{Muller2001,Muller2005,Endres2016}. These entries include the first three (CH$_3$OH) and two ($^{13}$CH$_3$OH and CH$_3^{18}$OH) torsional states and are based on the works of \citet{Xu2008}, \citet{Xu1997}, and \citet{Fisher2007}, respectively. The difference between the statistical weight factors $g_\mathrm{I}$ of $^{13}$CH$_3$OH ($g_\mathrm{I}=1$) and CH$_3^{18}$OH and CH$_3$OH ($g_\mathrm{I}=4$) is correctly taken into account in the CDMS database entries and therefore does not affect any column densities derived in this work. The line list of CH$_2$DOH is taken from the JPL catalog\footnote{\url{https://spec.jpl.nasa.gov/}} \citep{Pickett1998}, where the entry is based on the work of \citet{Pearson2012}. The line list of CHD$_2$OH is taken from \citet{Drozdovskaya2022}, which is mostly based on the the work of \citet{Coudert2021}. 

Only transitions with $E_\mathrm{up} \geq 50$~K are used to derive the column densities since lines with lower $E_\mathrm{up}$ likely include also emission from non-thermally desorbed methanol and emission possibly related to outflows. For $^{13}$CH$_3$OH, the $14_{1,13}-13_{2,12}$ ($E_\mathrm{up} = 254$~K) transition gives the only constraint on the column density for many sources as the other transitions suffer from severe line blending. Similarly, for CH$_3^{18}$OH only the $8_{1,8}-7_{0,7}$ ($E_\mathrm{up} = 86$~K) and $14_{1,14}-13_{2,12}$ ($E_\mathrm{up} = 239$~K) transitions provide constraints on the column density as well as some information on the excitation temperature. 
Furthermore, for CH$_2$DOH the $5_{2,4}\,e_1-4_{1,5}\,e_1$ ($E_\mathrm{up} = 59$~K) line, as well as several other lines, have rather low Einstein $A_\mathrm{ij}$ values ($<10^{-5}$~s$^{-1}$) and are often blended with other COMs. Moreover, the spectroscopy of the CH$_2$DOH $18_{1,17}\,o_1-18_{2,17}\,e_0$ line is unreliable and shows large discrepancies in $A_\mathrm{ij}$ between the JPL catalog entry ($A_\mathrm{ij} = 1.8 \times 10^{-5}$~s$^{-1}$) and that derived by \citet[$A_\mathrm{ij} = 8.9 \times 10^{-7}$~s$^{-1}$]{Coudert2014} and is therefore also excluded from the analysis. Consequently, the $17_{1,16}\,e_0-17_{0,17}\,e_0$ ($E_\mathrm{up} = 336$~K) transition of CH$_2$DOH provided the best constraint on the column density of CH$_2$DOH. However, although the $5_{1,5}\,e_0-4_{1,4}\,e_0$ ($E_\mathrm{up} = 36$~K) transition is excluded from the fitting, it can provide information on the excitation temperature of CH$_2$DOH as the best-fit LTE model should not overproduce this line. Lastly, for CHD$_2$OH, the $7_{0,1}\,e_1-6_{1,1}\,e_1$ ($E_\mathrm{up} = 74$~K) transition is the only detected line in our sample and therefore is the only constraint on the column density of CHD$_2$OH.

\begin{figure*}[t]
\centering
\includegraphics[height=4.77cm]{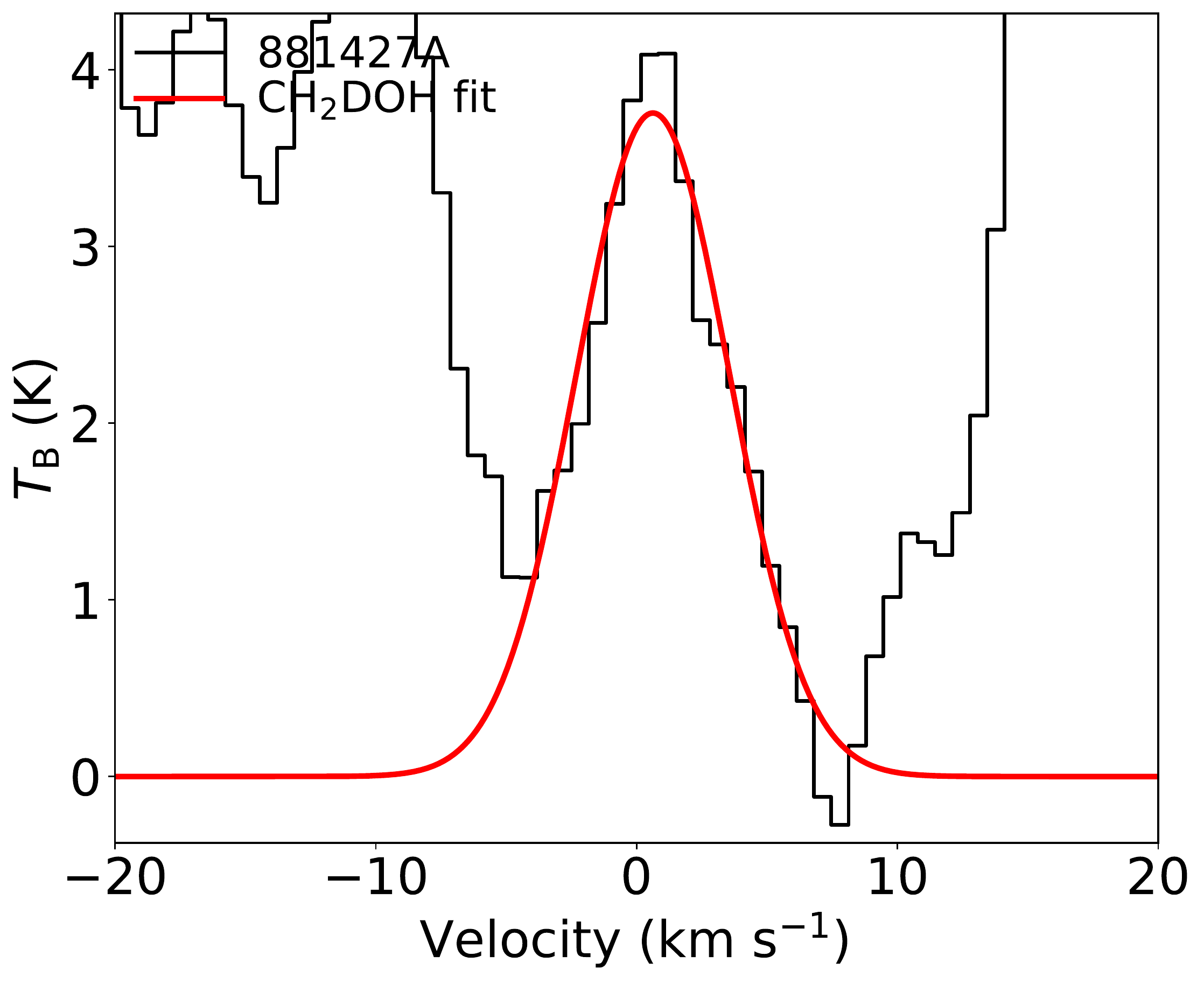}
\includegraphics[height=4.77cm]{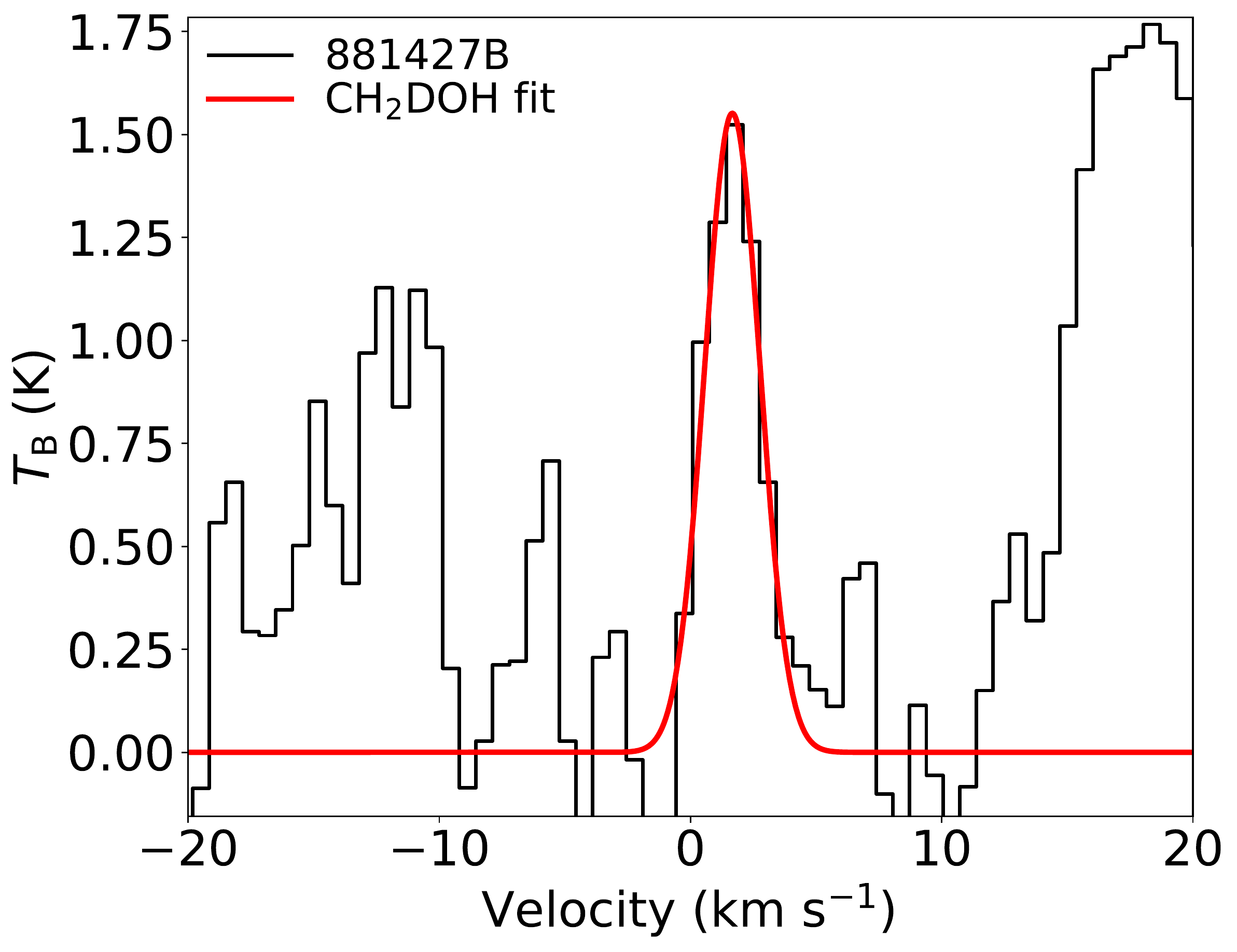}
\includegraphics[height=4.77cm]{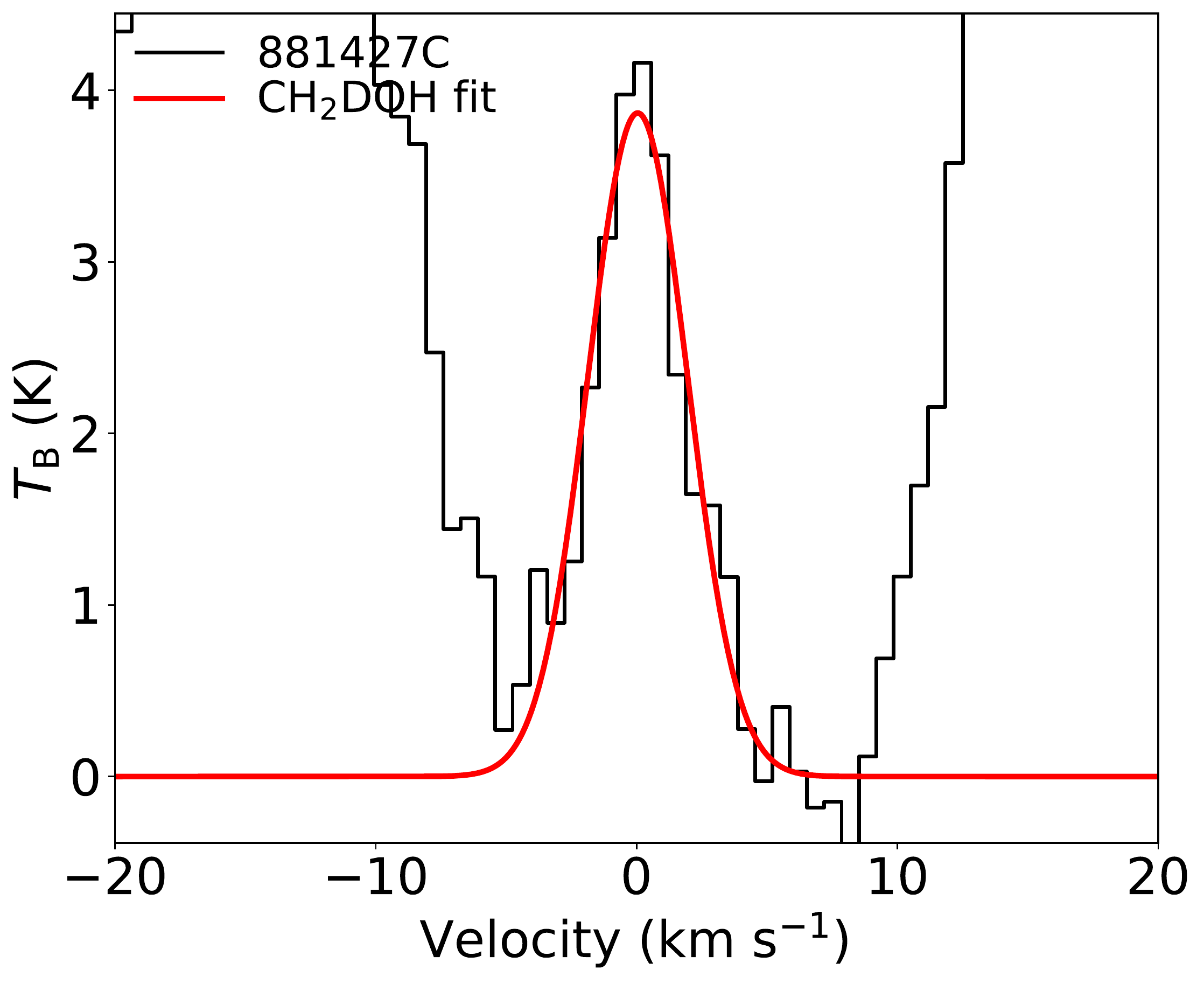}
\includegraphics[height=4.77cm]{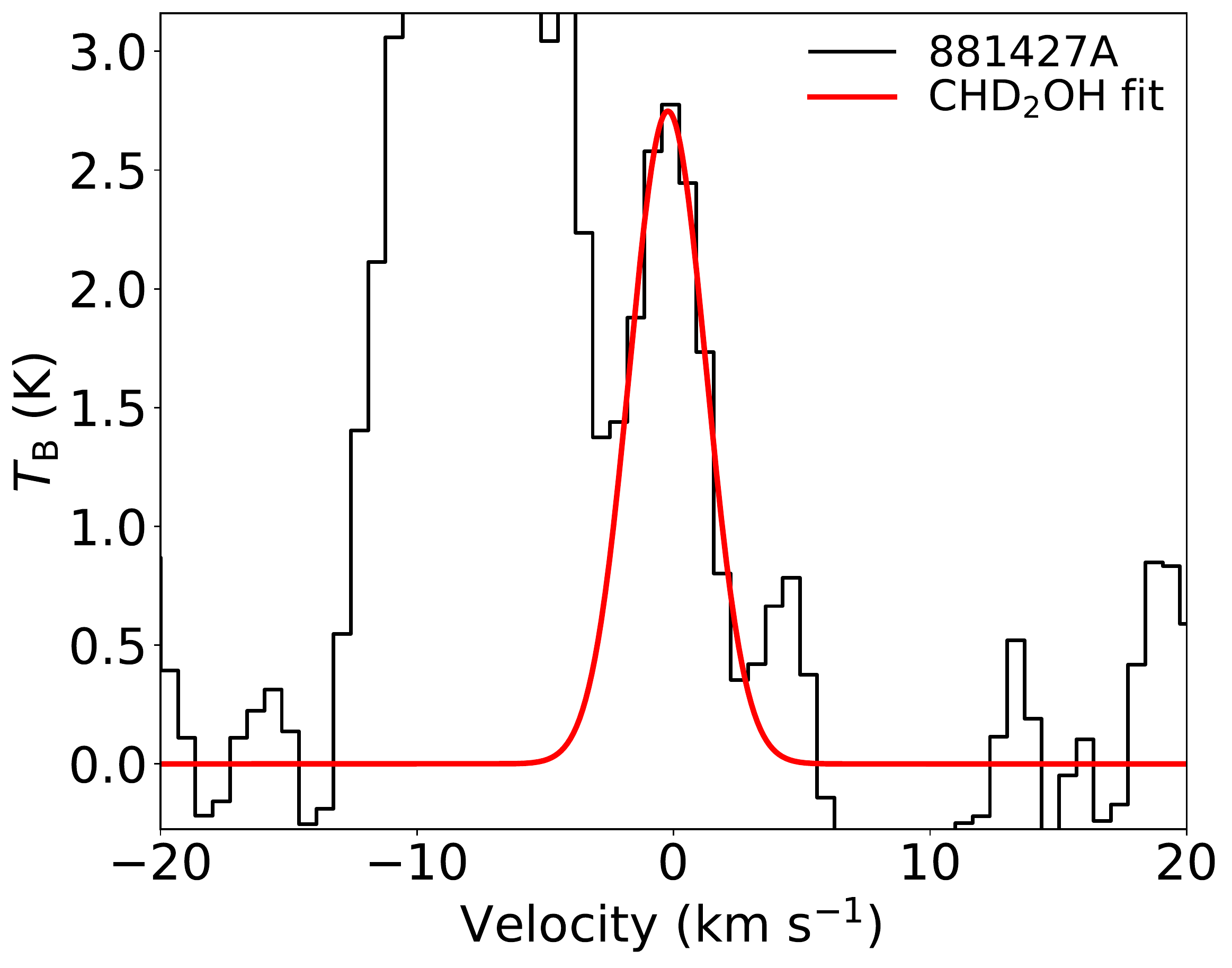}
\includegraphics[height=4.77cm]{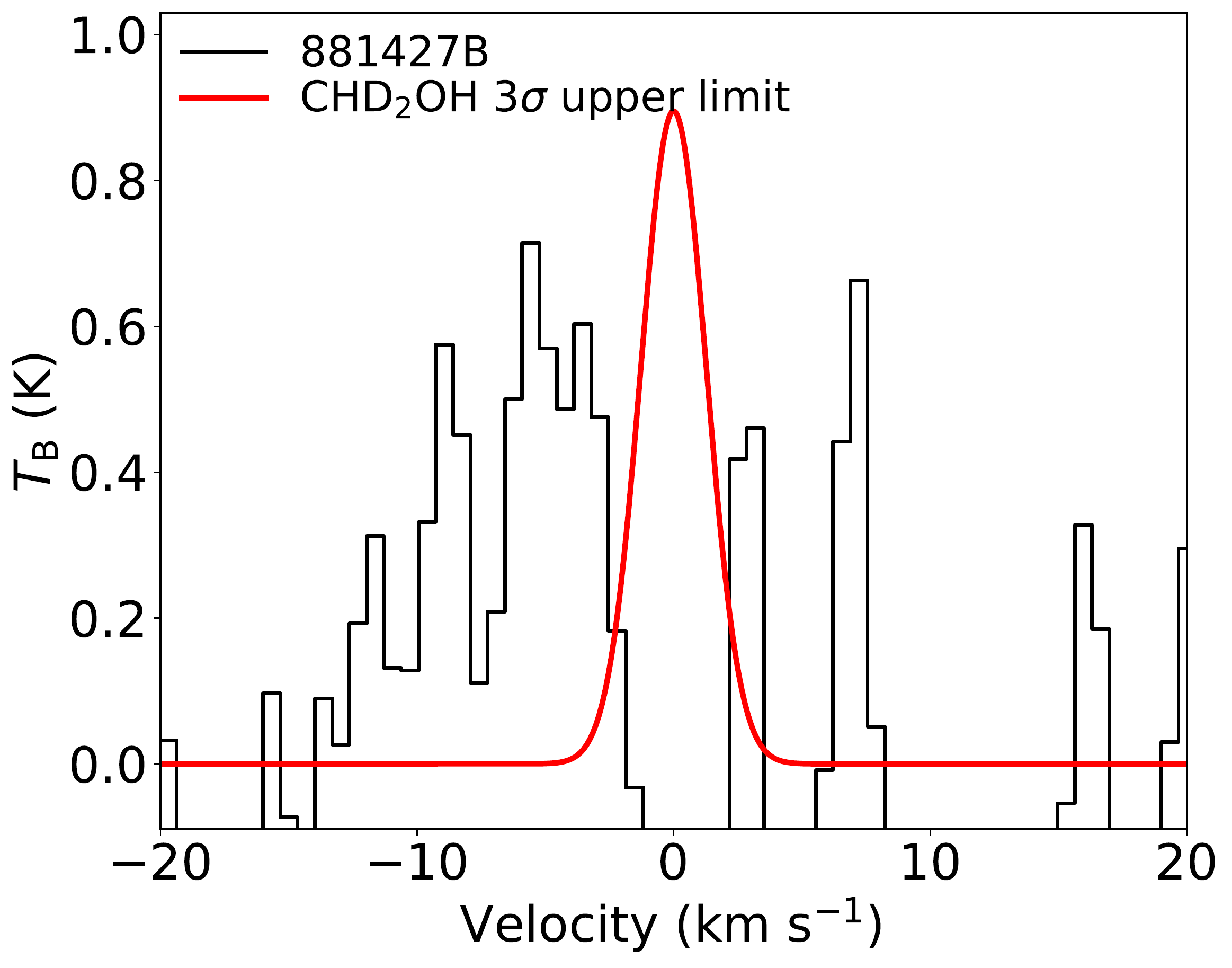}
\includegraphics[height=4.77cm]{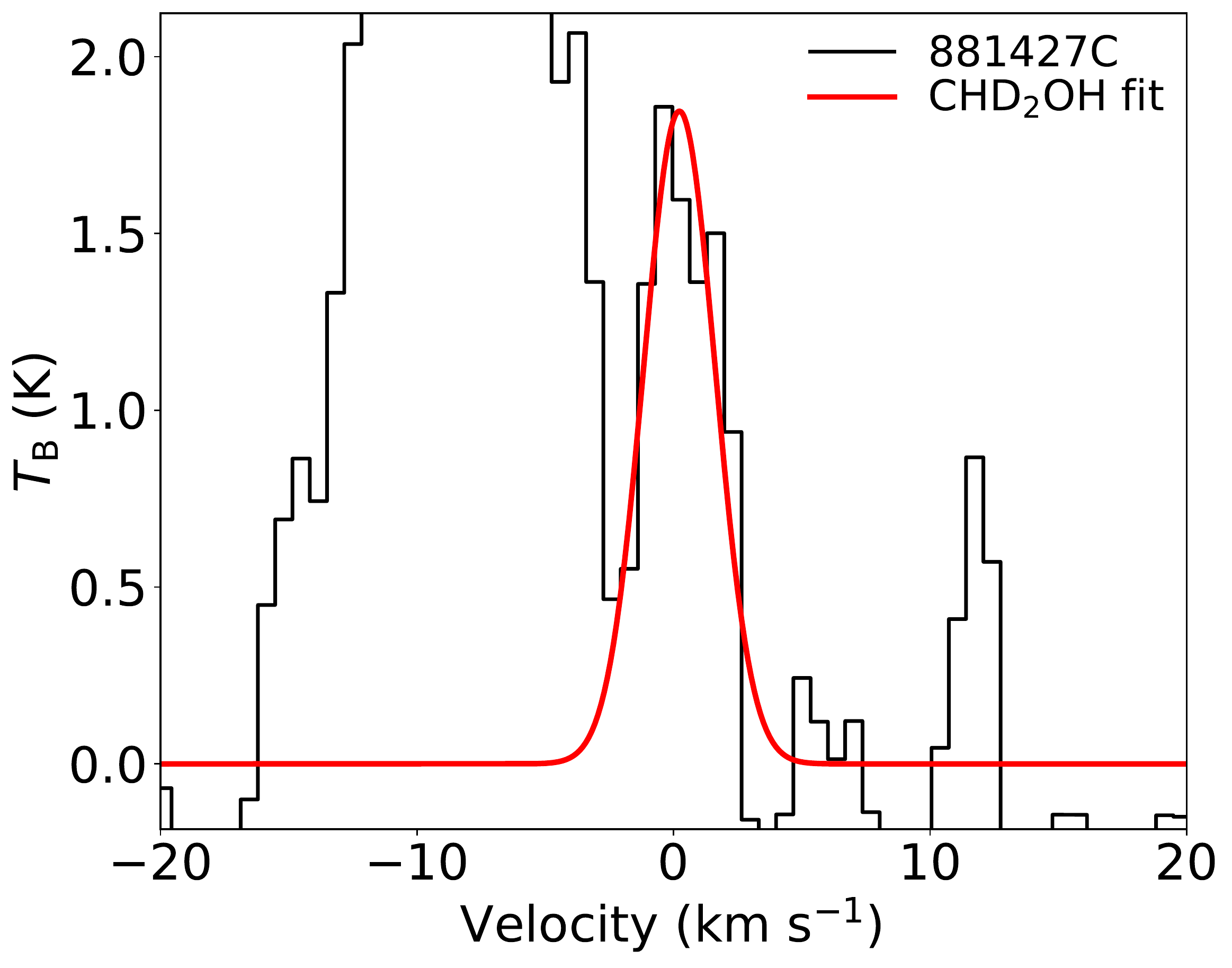}
\caption{Spectral line fits of CH$_2$DOH $17_{1,16}\,e_0-17_{0,17}\,e_0$ ($E_\mathrm{up} = 336$~K, top row) and CHD$_2$OH $7_{0,1}\,e_1-6_{1,1}\,e_1$ ($E_\mathrm{up} = 74$~K, bottom row) for 881427A (left), 881427B (middle), and 881427C (right). The data corrected for the $V_\mathrm{lsr}$ are shown in black and the best fit for $T_\mathrm{ex}=150$~K is shown in red. }
\label{fig:linefits_881427}
\end{figure*}

As a consequence of only single or a few lines being available, the excitation temperature is fixed to 150~K, which is roughly the mean temperature as measured toward other high-mass hot cores \citep[e.g.,][]{Neill2013,Belloche2016,Bogelund2018,Bogelund2019}. However, if clear anticorrelations between the best-fit LTE model and the data were present, the excitation temperature was varied by eye in steps of 25~K until the anticorrelations disappear in a similar way as the by eye fit method of \citet[][see their Appendix~C]{Nazari2021}.   

The column densities $N$ of $^{13}$CH$_3$OH, CH$_3^{18}$OH, CH$_2$DOH, and CHD$_2$OH are derived following a similar method as \citet{vanGelder2020}. A grid of $N$ and the full width at half maximum (FWHM) of the line is set and a model spectrum is computed for each grid point assuming LTE conditions. The size of the emitting region is fixed to the size of the beam (see Appendix~\ref{app:Observational_details}). Blended lines are excluded from the fitting procedure and similarly broad lines ($\mathrm{FWHM}\geq10$~\kms) and lines with $E_\mathrm{up} \leq 50$~K are excluded in the fit to exclude any emission possibly related to outflows. The best-fit column density and the $2\sigma$ uncertainty are computed from the grid for each isotopologue. 
The main contributors to the uncertainty of $N$ are the uncertainty on the flux calibration of ALMA (assumed to be 10\%) and the assumed excitation temperature. However, changing the excitation temperature in the 100-300~K range leads to at most a factor 3 variation in the derived column densities.
For several sources (e.g., 705768), the lines are broad ($>7$~km~s$^{-1}$) making automated line fitting complicated. For these sources the column density is estimated using the by eye fitting method of \citet{Nazari2021}. In this case, a 50\% uncertainty on the column density is assumed. Moreover, the sources 101899, 615590, 865468, and G345.5043+00.3480 showed line profiles consisting of multiple components. The column density of each component is derived and reported separately. 

For all sources, the column density of CH$_3$OH is derived from CH$_3^{18}$OH and, when no lines originating from CH$_3^{18}$OH were detected, from $^{13}$CH$_3$OH. The adopted $\mathrm{^{12}C/^{13}C}$ and $\mathrm{^{16}O/^{18}O}$ ratios are dependent on the galactocentric distance and are determined using the relations of \citet{Milam2005} and \citet{Wilson1994}, respectively. In cases where only upper limits on the column densities of both $^{13}$CH$_3$OH and CH$_3^{18}$OH could be derived, the range in $N_\mathrm{CH_3OH}$ was calculated by setting the 3$\sigma$ upper limit based on scaling the 3$\sigma$ upper limit of $^{13}$CH$_3$OH and the lower limit based on the main isotopologue. Lastly, when CH$_3$OH was not detected, the 3$\sigma$ upper limit is derived directly from CH$_3$OH lines.



\section{Results}
\label{sec:results}
The derived column densities of all isotopologues are presented in Table~\ref{tab:D_H_columns} for the reported excitation temperature. In Fig.~\ref{fig:linefits_881427}, the best-fit models to the CH$_2$DOH $17_{1,16}\,e_0-17_{0,17}\,e_0$ and CHD$_2$OH $7_{0,1}\,e_1-6_{1,1}\,e_1$ lines are presented for three hot cores in 881427 (see Fig.~\ref{fig:mom0_881427}).
Toward 25 sources, at least one clean unblended line of CH$_2$DOH is detected at the $3\sigma$ level, allowing for the determination of the column density. For the remaining 74 sources where no (unblended) transitions of CH$_2$DOH are detected, the $3\sigma$ upper limit is reported. For CHD$_2$OH, the column density could be determined for 11 sources.
In Table~\ref{tab:D_H_columns}, the column densities of $^{13}$CH$_3$OH, CH$_3^{18}$OH, and CH$_3$OH are also reported. 

\begin{figure*}
\includegraphics[width=\linewidth]{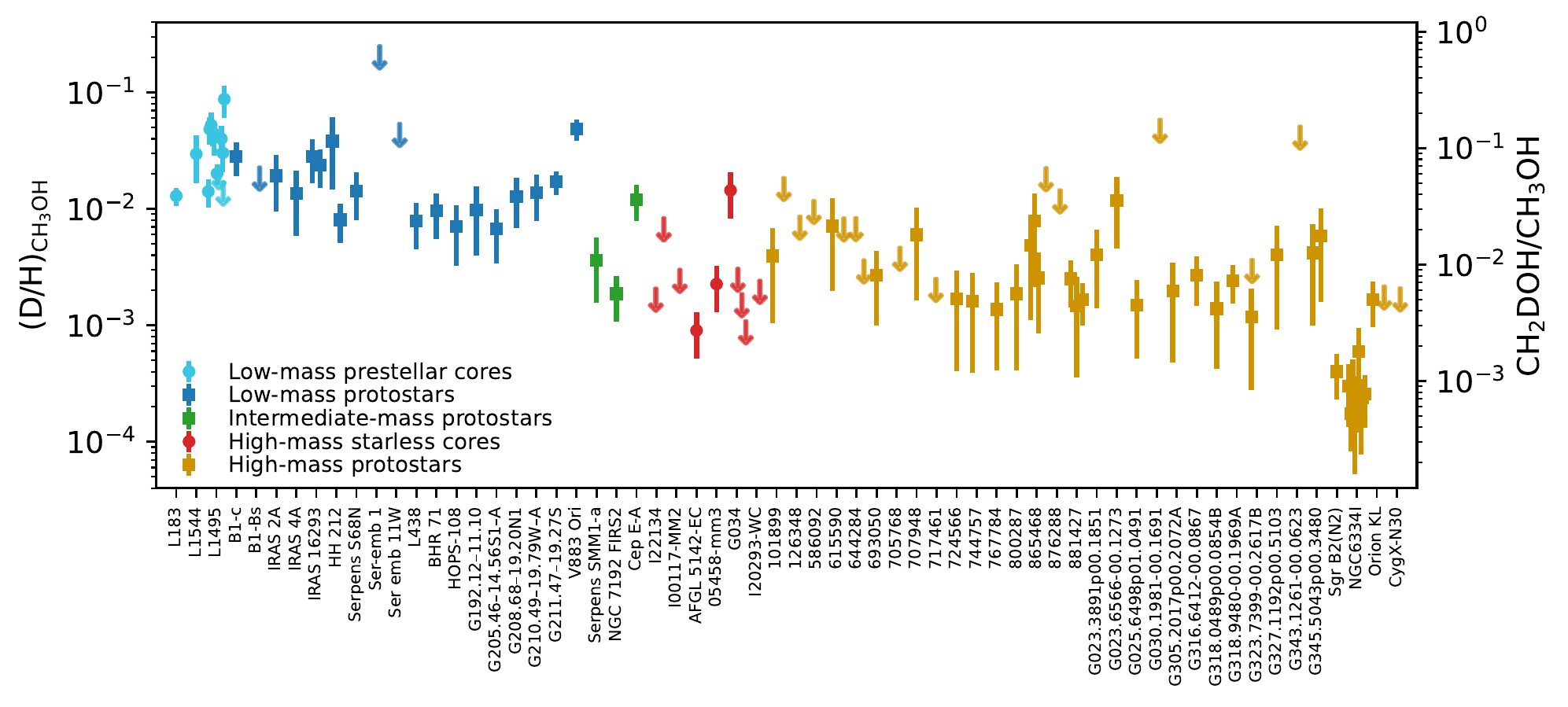}
\caption{The $\rm (D/H)_{CH_3OH}$ ratio (squares) derived from the $N_\mathrm{CH_2DOH}/N_\mathrm{CH_3OH}$ for low-mass, intermediate-mass, and high-mass protostellar systems including data from both this study and the literature (see Appendix~\ref{app:D_H_literature} for references). Only D/H ratios derived from interferometric observations are included to minimize effects of beam dilution and to exclude any contribution from larger scales. Also data for low-mass prestellar cores and high-mass starless cores (circles) from the literature (including observations with single dish telescopes) are included. Upper limits are presented as arrows.}
\label{fig:CH2DOH_CH3OH_D_H_names}
\end{figure*}

\begin{figure*}
\includegraphics[width=\linewidth]{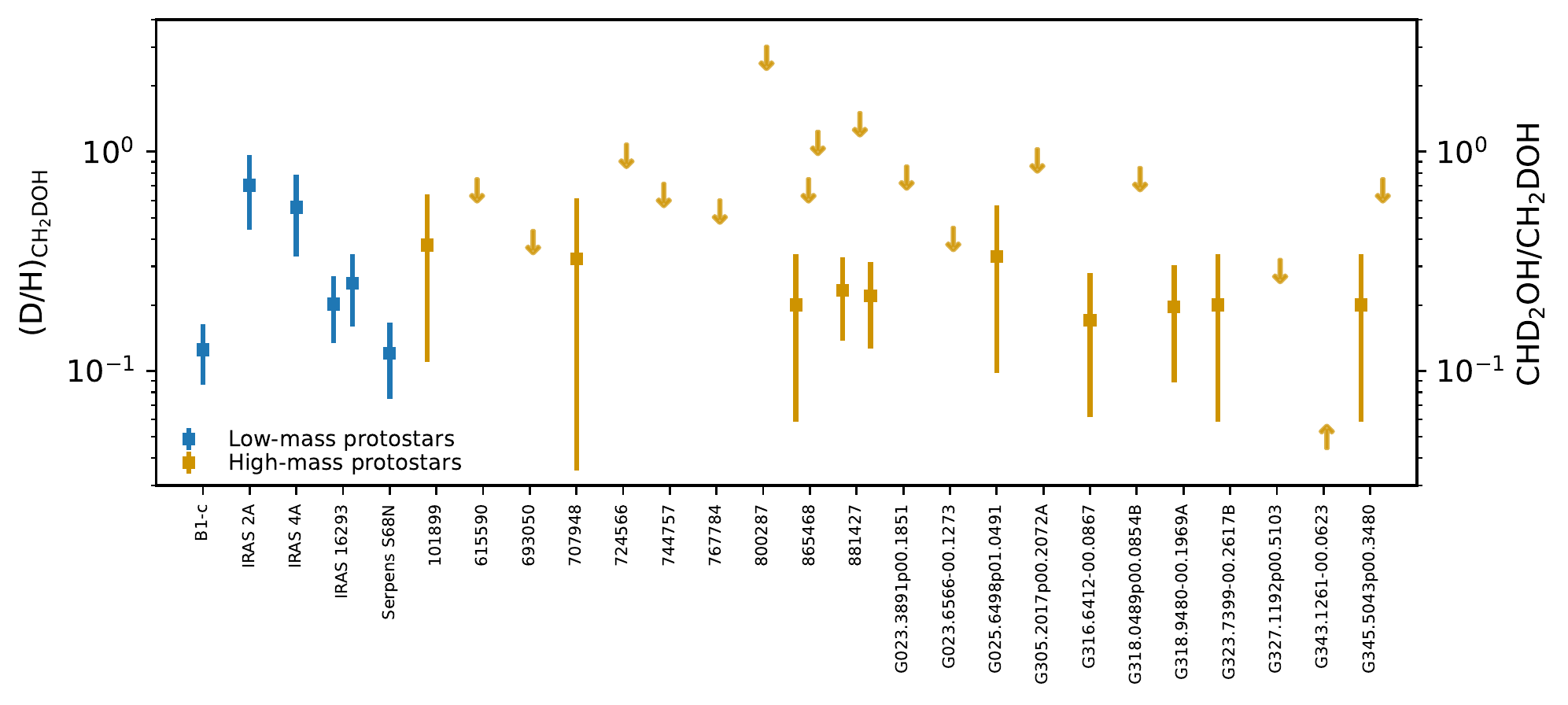}
\caption{The $\rm (D/H)_{CH_2DOH}$ ratio derived from the $N_\mathrm{CHD_2OH}/N_\mathrm{CH_2DOH}$ for low-mass and high-mass protostellar systems including data from both this study and the literature (see Appendix~\ref{app:D_H_literature} for references). Upper and lower limits are presented as arrows. Only D/H ratios derived from interferometric observations are included to minimize effects of beam dilution and to exclude any contribution from larger scales. The $\rm (D/H)_{CH_2DOH}$ of IRAS~2A and IRAS~4A were derived using older spectroscopic data \citep{Taquet2019}.}
\label{fig:CHD2OH_CH2DOH_D_H_names}
\end{figure*}

The column densities of CH$_2$DOH are generally between one and three orders of magnitude lower than those of CH$_3$OH, see also Fig.~\ref{fig:CH2DOH_CH3OH_D_H_names}. Furthermore, the column densities of CHD$_2$OH are about a factor 3--10 lower than that of CH$_2$DOH, see Fig.~\ref{fig:CHD2OH_CH2DOH_D_H_names}. In order to translate the column density ratios to the D/H ratios, statistical weighting has to be taken into account since a deuterium atom has a three times higher probability to land in the CH$_3$ group compared to the OH group. Therefore, the D/H ratios of CH$_3$OH and CH$_2$DOH can be derived through,
\begin{align}
N_{\rm CH_2DOH}/N_{\rm CH_3OH} & = \rm 3 (D/H)_{CH_3OH}, \label{eq:CH3OH_D_H_CH2DOH} \\
N_{\rm CHD_2OH}/N_{\rm CH_2DOH} & = \rm (D/H)_{CH_2DOH}. \label{eq:CH2DOH_D_H_CHD2OH}
\end{align}
The derived D/H ratios are also listed in Table~\ref{tab:D_H_columns}.

The resulting $\rm (D/H)_{CH_3OH}$ and $\rm (D/H)_{CH_2DOH}$ ratios are presented in Figs.~\ref{fig:CH2DOH_CH3OH_D_H_names} and \ref{fig:CHD2OH_CH2DOH_D_H_names}, respectively. Including upper limits, a (limit on the) $\rm (D/H)_{CH_3OH}$ and $\rm (D/H)_{CH_2DOH}$ ratios could be derived for 38 and 26 of the 99 studied sources, respectively. 
Besides the ALMAGAL sources, also other classical high-mass hot cores such as Sgr~B2(N2) \citep{Belloche2016}, NGC~6334I \citep{Bogelund2018}, Orion~KL \citep{Neill2013}, and CygX-N30 \citep{vanderWalt2021} are included in Figs.~\ref{fig:CH2DOH_CH3OH_D_H_names} and \ref{fig:CHD2OH_CH2DOH_D_H_names}. Only sources where $N_\mathrm{CH_3OH}$ is derived from the $^{13}$C or $^{18}$O isotopologues are included in Fig.~\ref{fig:CH2DOH_CH3OH_D_H_names} to ensure that $N_{\rm CH_3OH}$ is not underestimated. 
The $\rm (D/H)_{CH_3OH}$ ratios lie mostly in the $10^{-2}-10^{-4}$ range. 
Interestingly, all the ALMAGAL sources and Orion~KL show higher $\rm (D/H)_{CH_3OH}$ ratios ($10^{-2}-10^{-3}$) than Sgr~B2(N2) and NGC~6334I ($10^{-3}-10^{-4}$). 
No clear correlation between the detection of CH$_2$OH or the derived $\rm (D/H)_{CH_3OH}$ and protostellar parameters such as $L_\mathrm{bol}$ and envelope mass is present among the high-mass sources.
Excluding upper limits, the average $\rm (D/H)_{CH_3OH}$ ratio is $1.1\pm0.7\times10^{-3}$. This is almost two orders of magnitude higher than the D/H ratio in the local ISM of $\sim2\times10^{-5}$ \citep{Linsky2006,Prodanovic2010}, suggesting effective deuteration in the cold high-mass prestellar phases. However, both the range of observed $\rm (D/H)_{CH_3OH}$ values and the average is more than one order of magnitude lower than what is generally observed toward low-mass sources \citep[$\sim\mathrm{few}\times10^{-2}$, e.g.,][see Sect.~\ref{subsec:D/H_lowmass_highmass} for further discussion]{Bianchi2017_SVS13A,Bianchi2017_HH212,Bianchi2020,Jacobsen2019,vanGelder2020}.

Interestingly, the $\rm (D/H)_{CH_2DOH}$ ratio (Eq.~\eqref{eq:CH2DOH_D_H_CHD2OH}) is significantly higher than the $\rm (D/H)_{CH_3OH}$ ratio, see Fig.~\ref{fig:CHD2OH_CH2DOH_D_H_names}. For the high-mass sources, only ALMAGAL datapoints are shown since no other interferometric studies of CHD$_2$OH in high-mass protostellar systems are available. The derived $\rm (D/H)_{CH_2DOH}$ ratios lie mostly in the $0.1-1$ range, with an average of $2.0\pm0.8\times10^{-1}$, which is more than two orders of magnitude higher than the $\rm (D/H)_{CH_3OH}$ ratio. Furthermore, this indicates that about 1/5 of the single deuterated methanol molecules gets successively deuterated further toward CH$_2$DOH in high-mass protostellar systems. This is in good agreement with low-mass protostellar systems where about 1/4 of the CH$_2$DOH is successively deuterated toward CHD$_2$OH \citep{Drozdovskaya2022}.





\section{Discussion}
\label{sec:discussion}
\subsection{Methanol deuteration from low to high mass}
\label{subsec:D/H_lowmass_highmass}
In this work, a (limit on the) $\rm (D/H)_{CH_3OH}$ ratio could be derived for 38 of the 99 studied high-mass sources. Since large samples of both low-mass and high-mass protostellar systems with methanol D/H values are now available, a more significant comparison over the mass regime can be made. 
In Fig.~\ref{fig:CH2DOH_CH3OH_D_H_names}, also $\rm (D/H)_{CH_3OH}$ ratios derived for both low-mass prestellar cores and high-mass starless cores are included. 
It is evident that the $\rm (D/H)_{CH_3OH}$ ratio is lower in high-mass hot cores ($10^{-4}-10^{-2}$) than in their low-mass counterpart ($10^{-2}-10^{-1}$). Intermediate-mass protostars show values in between ($10^{-3}-10^{-2}$), but this subsample only consists of three sources \citep[NGC~7192~FIR2, Cep~E-A, and Serpens~SMM1-a,][the D/H ratio of NGC~7192~FIR2 is taken  from the beam averaged values]{Fuente2014,Ospina-Zamudio2018,Ligterink2022}. However, interestingly the $\rm (D/H)_{CH_2DOH}$ ratio seems very similar between low-mass protostars and high-mass protostars ($0.1-1$, see Fig.~\ref{fig:CHD2OH_CH2DOH_D_H_names}). Among the low-mass sources, IRAS~2A and IRAS~4A show somewhat elevated $\rm (D/H)_{CH_2DOH}$, but these were derived using older spectroscopic data of CHD$_2$OH \citep{Taquet2019}.

\begin{figure}
\includegraphics[width=\linewidth]{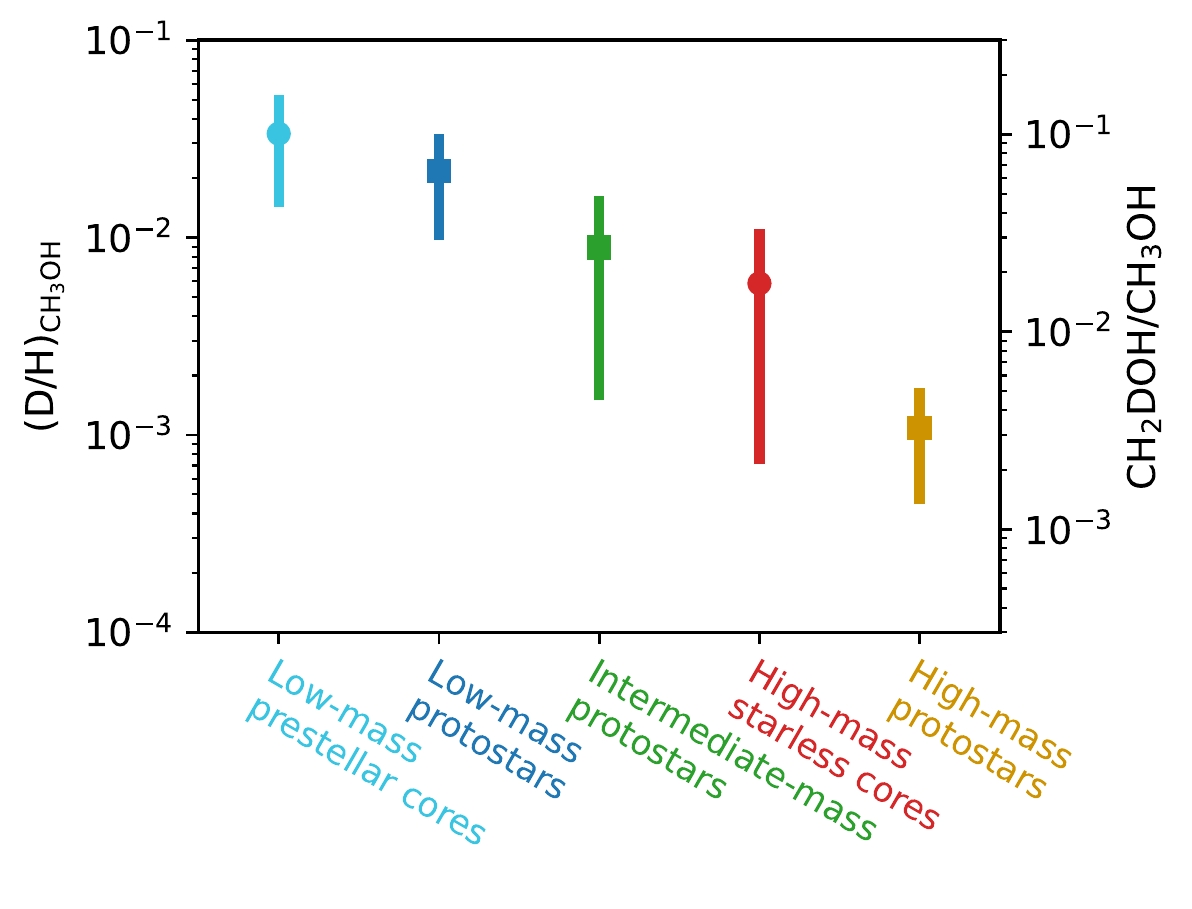}
\caption{
The average $\rm (D/H)_{CH_3OH}$ ratio derived from $N_\mathrm{CH_2DOH}/N_\mathrm{CH_3OH}$ 
for low-mass, intermediate-mass, and high-mass protostellar systems and low-mass prestellar cores and high-mass starless cores. The prestellar phases are indicated with circles and the protostellar phases with squares.
}
\label{fig:D_H_averages}
\end{figure}

In Fig.~\ref{fig:D_H_averages}, the mean $\rm (D/H)_{CH_3OH}$ ratio is presented for low-mass, intermediate-mass, and high-mass protostellar systems. The average $\rm (D/H)_{CH_3OH}$ ratio for high-mass hot cores ($1.1\pm0.7\times10^{-3}$) lies more than one order of magnitude lower than the average ratio for low-mass hot corinos ($2.2\pm1.2\times10^{-2}$), with the average $\rm (D/H)_{CH_3OH}$ for intermediate-mass protostars in between ($8.9\pm7.4\times10^{-3}$). A similar trend is seen for the high-mass and low-mass prestellar phases where the average $\rm (D/H)_{CH_3OH}$ ratios are $5.9\pm5.1\times10^{-3}$ and $3.4\pm1.9\times10^{-2}$, respectively. The lower $\rm (D/H)_{CH_3OH}$ ratio in both high-mass hot cores and high-mass prestellar phases compared to their lower-mass counterparts suggests a lower deuteration efficiency already in the high-mass prestellar phases (see Sect.~\ref{subsec:D_H_GRAINOBLE}). 


The methanol D/H ratios derived in low-mass prestellar cores agree well with those derived for low-mass protostars (see Fig.~\ref{fig:CH2DOH_CH3OH_D_H_names}). Since CH$_3$OH is formed through the hydrogenation of CO ice \citep[e.g.,][]{Watanabe2002,Fuchs2009,Simons2020,Santos2022}, this is a strong indication for inheritance of methanol and other COMs ices between low-mass prestellar phases and protostars. On the other hand, the average $\rm (D/H)_{CH_3OH}$ ratio for high-mass starless cores ($5.9\pm5.1\times10^{-3}$) seems to be about a factor of five higher than that for high-mass protostars ($1.1\pm0.7\times10^{-3}$). However, the average $\rm (D/H)_{CH_3OH}$ of the high-mass starless cores has a large errorbar since it is based on three detections of which one (G034-G2(MM2)) has a high $\rm (D/H)_{CH_3OH}$ ratio of $\sim10^{-2}$ \citep{Fontani2015}. The other two detections \citep[AFGL~5142-EC and 0548-mm3;][]{Fontani2015} and all the upper limits show $\mathrm{D/H} \lesssim2\times10^{-3}$ which agree well with most of the ALMAGAL sources as well as with Orion~KL \citep{Neill2013}. Only Sgr~B2(N2) and most of the cores in NGC~6334I show slightly lower D/H ratios at the $10^{-4}$ level. This therefore also suggests inheritance of methanol ice from the high-mass prestellar phase to the protostellar phase. 

\subsection{Singly vs doubly deuterated methanol}
\label{subsec:CH2DOH_CHD2OH_discrepancy}
As evident from Figs.~\ref{fig:CH2DOH_CH3OH_D_H_names} and \ref{fig:CHD2OH_CH2DOH_D_H_names}, the methanol D/H ratio derived for CH$_2$DOH is significantly higher than that derived for CH$_3$OH. The average $\rm (D/H)_{CH_2DOH}$ ratio is about two orders of magnitude higher ($2.0\pm0.8\times10^{-1}$) than $\rm (D/H)_{CH_3OH}$. In contrast to $\rm (D/H)_{CH_3OH}$, this is in good agreement with the average of $3.0\pm2.0\times10^{-1}$ for low-mass protostars, suggesting that successive deuteration happens almost equally effective in both low-mass and high-mass systems. 

Having higher D/H ratios for the doubly deuterated isotopologue compared with singly deuterated isotopologue is not unique to methanol. For water, the D$_2$O/HDO ratios are on the order of $10^{-2}$ \citep[e.g.,][]{Coutens2014,Jensen2021}, which is about an order of magnitude higher than typical HDO/H$_2$O ratios \citep[$\lesssim10^{-3}$;][]{Persson2014,Jensen2019,vantHoff2022}. This difference was attributed to be the result of layered ice chemistry \citep{Dartois2003,Furuya2016}, where the bulk of the water ice is formed in the warmer translucent cloud phase with a low D/H ratio whereas the surface layers formed in the cold prestellar phases show higher D/H ratios. However, methanol is thought to only start forming in the cold prestellar phases where CO is frozen out \citep[e.g.,][]{Watanabe2002,Fuchs2009} with little to no formation in the warmer translucent phases. 
Indeed, also for a direct precursor of CH$_3$OH on the surface of dust grains, H$_2$CO, the D$_2$CO/HDCO ratio in IRAS~16293-2422 points toward a high D/H ratio of $\sim25$\% compared to a much lower D/H ratio derived from HDCO/H$_2$CO \citep[$\sim3$\%;][]{Persson2018}. 
Small variations in temperature in the 10--20~K range can change the D/H ratio of ice mantle species such as methanol (see Sect.~\ref{subsec:D_H_GRAINOBLE}), but this should affect both CH$_2$DOH and CHD$_2$OH in a similar way and should therefore not lead to the observed difference. 

One possible explanation could be the optical depth of CH$_2$DOH. In the low-mass source L1551~IRS5, the emission of CH$_2$DOH (as well as $^{13}$CH$_3$OH) was suggested to be optically thick \citep{Bianchi2020}. However, since the $\rm (D/H)_{CH_3OH}$ ratios derived from CH$_2$DOH clearly show lower values in high-mass protostellar systems compared to their lower-mass counterpart (see Sect.~\ref{subsec:D/H_lowmass_highmass}), this does not seem like a viable solution. Very recently, spectroscopic data for $^{13}$CH$_2$DOH has become available \citep{Ohno2022}, but these do not yet include a calculation of the partition function and line properties such as $A_\mathrm{ij}$. When assuming that the source size is equal to the beam size, the line optical depth of the most constraining transition, ($17_{1,16}\,e_0-17_{0,17}\,e_0$, $E_\mathrm{up} = 336$~K) is $\tau<10^{-2}$. Only for source sizes smaller than $<0.5''$ does CH$_2$DOH become marginally optically thick ($\tau>0.1$) for the most line rich sources. Also, the $\rm (D/H)_{CH_3OH}$ ratios where $N_\mathrm{CH_3OH}$ was derived from the possibly optically thick $^{13}$C isotopologue are on average less than a factor $\sim3$ higher than the $\rm (D/H)_{CH_3OH}$ ratios where $N_\mathrm{CH_3OH}$ could be derived using the optically thin $^{18}$O isotopologue (see Fig.~\ref{fig:CH2DOH_CH3OH_D_H_ALMAGAL_isotopologues}).

\begin{figure*}
\includegraphics[width=\linewidth]{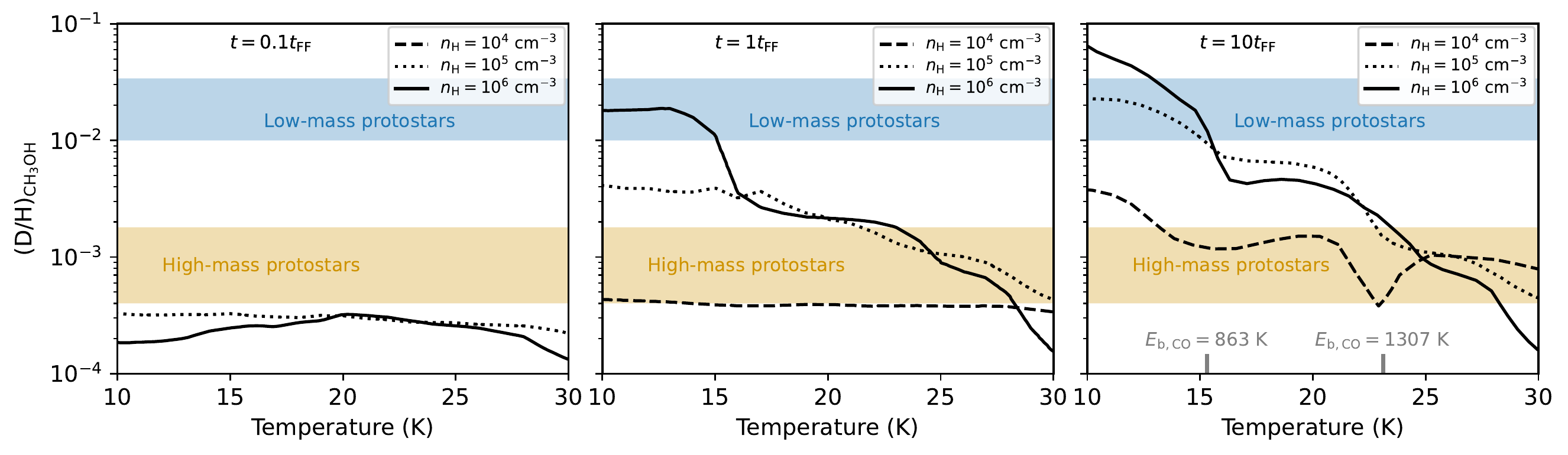}
\caption{The $\rm (D/H)_{CH_3OH}$ ratio in the ices as function of the gas and dust temperature for $n_\mathrm{H} = 10^{4}$ (dashed), $10^{5}$ (dotted), and $10^{6}$~cm$^{-3}$ (solid) as modeled by \citet{Bogelund2018} and \citet{Taquet2019} using the GRAINOBLE model \citep{Taquet2012,Taquet2013,Taquet2014}. The model results are shown for 0.1 (left), 1 (middle), and 10 (right) times free-fall timescales $t_\mathrm{FF}$. The observed average $\rm (D/H)_{CH_3OH}$ ratios are indicated in blue and orange for the low-mass and high-mass protostars, respectively. 
In the right panel, the range of possible desorption temperature of CO ice is indicated with the gray bars for binding energies between 863~K and 1307~K for CO ice deposited on non-porous amorphous solid water
\citep{Noble2012}.
}
\label{fig:D_H_models_HMP,LMP}
\end{figure*}

A more realistic explanation is that successive deuteration of molecules is more effective than the first deuteration. This explanation is supported by several laboratory studies performed at low temperatures of 10--20~K \citep[e.g.,][]{Nagaoka2005,Nagaoka2007,Hidaka2009}. \citet{Drozdovskaya2022} showed that their observed $\rm (D/H)_{CH_3OH}$ ratio as derived from CH$_2$DOH for the low-mass binary IRAS~16293-2422 could be well explained by these experiments whereas CHD$_2$OH and CD$_3$OH were overproduced by the experiments. The latter could be the result of the high atomic D/H flux of 0.1 used in the laboratory studies in contrast to the ISM value of $\sim10^{-5}$, although the atomic D/H ratio is enhanced in cold dense prestellar cores. Assuming that the $\rm (D/H)_{CH_3OH}$ ratio is a direct representative of the gaseous atomic D/H ratio available in the prestellar phases (i.e., that H/D addition reactions are equally effective), a D/H flux of $\sim10^{-3}$ may be more realistic for high-mass cold dense cores.

\subsection{Linking the methanol D/H to the physical conditions during formation}
\label{subsec:D_H_GRAINOBLE}
Given the sensitivity of the methanol deuteration process to both temperature and density (i.e., CO freeze-out), the measured methanol D/H ratios are linked to these physical properties during the prestellar phases. To quantify this for the protostellar systems studied in this work, the observed D/H ratios are compared to the astrochemical gas-grain models presented by \citet{Bogelund2018} and \citet{Taquet2019}. These works used the GRAINOBLE model \citep{Taquet2012,Taquet2013,Taquet2014} to test the effect of the dust and gas temperature $T$ (assumed to be equal) and hydrogen density $n_\mathrm{H} = n(\mathrm{H}) + 2 n(\mathrm{H_2})$ on the resulting $\rm (D/H)_{CH_3OH}$ ratio in the ices during the prestellar phases. In this work, we compare our results to their results and therefore only a brief description of the model is presented.

In GRAINOBLE, the gas-ice chemistry is computed in three phases: the bulk ice, the ice surface layers, and in the gas phase, following the approach initially presented by \citet{Hasegawa1993}. The model includes both adsorption and desorption reactions and computes the rate equations in each phase. The chemical network used for the gas-phase chemistry is described in \citet{Taquet2014} and includes both ion-neutral chemistry and all molecules relevant for the chemistry of methanol (e.g., CO, HCO, H$_2$CO). Moreover, the model computes the deuteration of ice species based on laboratory experiments and includes both the hydrogenation (with both H and D atoms) reactions leading to methanol as well as hydrogen/deuterium abstraction reactions in low temperature ($\sim10-15$~K) conditions \citep{Hidaka2009}.

The effect of $T$ and $n_\mathrm{H}$ on the resulting methanol D/H ratio is presented in Fig.~\ref{fig:D_H_models_HMP,LMP} as computed by \citet{Bogelund2018} and \citet{Taquet2019}. For a constant temperature and density, the chemistry is evolved over a timescale indicated on the top of each panel, where the free-fall timescale $t_\mathrm{FF}$ is $4.4\times10^{5}$, $1.4\times10^{5}$, and $4.4\times10^{4}$ years for $n_\mathrm{H} = 10^{4}$, $10^{5}$, and $10^{6}$~cm$^{-3}$, respectively. 
For the longest timescales ($t = 10 t_\mathrm{FF}$), it is evident that for all densities the methanol D/H ratio drops with increasing temperature. The strongest decrease is seen for $10^{6}$~cm$^{-3}$, where the methanol D/H ratio decreases from $\sim6\times10^{-2}$ for $T=10$~K to as low as $\sim10^{-4}$ when $T=30$~K. A similar trend is visible for $10^{5}$~cm$^{-3}$ where the D/H ratio decreases from $\sim2\times10^{-2}$ for $T=10$~K to $\sim4\times10^{-4}$ at $T\sim30$~K. The decreasing D/H ratios for both these densities is the direct consequence of the decrease of atomic deuterium enhancement in Eq.~\eqref{eq:H2D+_formation} with increasing temperature. For $n_\mathrm{H} = 10^{4}$~cm$^{-3}$, the methanol D/H ratio also decreases with temperature, but only by a factor $\sim4$ between $10-30$~K.

Another interesting trend is that for decreasing timescales (i.e., moving from right to left in Fig.~\ref{fig:D_H_models_HMP,LMP}), the D/H ratio at a given temperature also drops for all densities. The strongest drops are seen for the higher density cases at low temperatures ($<15$~K) where the D/H ratio drops two orders of magnitude from $t = 10 t_\mathrm{FF}$ toward $t = 0.1 t_\mathrm{FF}$. This is the direct result of having less time where CO is frozen out and hence less time to deuterate ice species such as methanol. 
For higher temperatures ($T>20$~K), this effect is less evident since significantly less CO freezes out, although a slightly higher binding energy of CO \citep[up to $\sim1300$~K;][]{Noble2012} could result in CO frozen out till higher temperatures of $\sim25$~K and hence a higher deuteration efficiency also above 20~K.
However, even when CO does not freeze out, CO molecules can still land on the grain for a short period and react with H or D atoms toward HCO, H$_2$CO, and eventually (deuterated) methanol. This effect is most efficient for higher densities of $10^{5}-10^{6}$~cm$^{-3}$ and reduces when timescales smaller than the free-fall timescale are considered. For the lowest density of $10^{4}$~cm$^{-3}$, this effect is most evident since the CO freeze-out timescale is the highest and therefore the methanol deuteration is hampered the most.


Overplotted in Fig.~\ref{fig:D_H_models_HMP,LMP} are the observed methanol $\rm (D/H)_{CH_3OH}$ ratios for both low-mass and high-mass protostars. It is evident that the observed $\rm (D/H)_{CH_3OH}$ ratios suggest a different temperature during methanol formation or different prestellar phase lifetimes for low-mass and high-mass protostars. For high-mass protostars, a temperature of $>20$~K is needed when the density is larger than $10^{5}$~cm$^{-3}$ and the timescale of the high-mass prestellar phase is $>t_\mathrm{FF}$. A lower temperature of $>13$~K at low densities of $10^{4}$~cm$^{-3}$ can also explain the observed methanol D/H ratio toward high-mass protostars, but such low densities in the dense high-mass starless phase are unlikely. Alternatively, the temperature in the high-mass prestellar phases can be in the $15<T<20$~K range with a lifetime of $\lesssim t_\mathrm{FF}$. For prestellar lifetimes much smaller than the free-fall timescale, any temperature can explain the observed methanol deuteration toward high-mass protostars.

Contrary to the high-mass protostars, the observed $\rm (D/H)_{CH_3OH}$ ratio for low-mass protostars suggests both a temperature of $<15$~K and a prestellar phase duration longer than $\geq t_\mathrm{FF}$. Furthermore, the observed methanol D/H ratio for low-mass protostars can not be explained by a low density of $10^{4}$~cm$^{-3}$ at any modeled timescale. 

It is important to note that the methanol D/H ratio is observed in the gas phase with ALMA whereas the \textsc{GRAINOBLE} models predict the ice abundances in the prestellar phases. Several processes can affect the D/H ratio as the ices warm up while infalling toward the protostar \citep[e.g.,][]{Ratajczak2009,Faure2015}. However, one of the likely dominant processes, CH$_2$DOH formation through H-D substitution in methanol ice \citep{Nagaoka2005} is included in the model but this does not dominate over hydrogenation of CO. 

These results thus suggest that the high-mass prestellar phases are generally either warm ($T\gtrsim20$~K) or short ($t\lesssim t_\mathrm{FF}$) while the low-mass prestellar phases are colder ($T<15$~K) and long ($t\geq t_\mathrm{FF}$). The observed methanol D/H ratios toward high-mass starless cores and low-mass prestellar cores also fit this picture, see Fig.~\ref{fig:D_H_models_HMSC,LMPC}. 
On the other hand, the spread in observed abundance ratios of nitrogen-bearing COMs suggests that the scatter in timescales of high-mass prestellar phases is rather small and similar to that of low-mass prestellar phases \citep{Nazari2022_ALMAGAL}, implying that warmer high-mass pre-stellar phases are a more likely explanation.
One possible explanation for the discrepancy between the low-mass and high-mass methanol D/H ratios could be that high-mass stars generally form in clusters with other nearby high-mass stars that heat the surrounding cloud which can affect the D/H ratios molecules forming in the ices \citep[e.g., such as seen for water toward low-mass protostars;][]{Jensen2019}.
The majority of the sources studied in this work are located in a clustered environments but these do not show significantly lower D/H ratios than high-mass sources that are single sources at our angular resolution.
More modeling work similar to those performed by \citet{Bogelund2018} and \citet{Taquet2019} including CHD$_2$OH is needed to further test these hypotheses. 

\section{Conclusion}
\label{sec:conclusion}
In this work, (limits on) the D/H ratios of CH$_3$OH and CH$_2$DOH are determined for 38 and 26 sources, respectively, out of the 99 studied sources using ALMA observations of CH$_2$DOH, CHD$_2$OH, CH$_3$OH, $^{13}$CH$_3$OH, and CH$_3^{18}$OH. The derived $\rm (D/H)_{CH_3OH}$ and $\rm (D/H)_{CH_2DOH}$ ratios are compared to each other as well as to other high-mass protostars, low-mass protostars, and both low-mass and high-mass prestellar phases. Furthermore, comparison with the gas-grain chemical code GRAINOBLE links the observed D/H ratios to the temperature during methanol formation and the lifetime of the prestellar phases. The main conclusions of this work are as follows:
\begin{itemize}
\item 
The $\rm (D/H)_{CH_3OH}$ ratios of the high-mass protostars studied in this work lie in the $10^{-3}-10^{-2}$ range.
Combining our sample with other high-mass protostars studied with ALMA, an average $\rm (D/H)_{CH_3OH}$ ratio of $1.1\pm0.7\times10^{-3}$ is derived. This is in good agreement with the $\rm (D/H)_{CH_3OH}$ ratio derived for high-mass starless cores ($5.9\pm5.1\times10^{-3}$), but is more than an order of magnitude lower than the average $\rm (D/H)_{CH_3OH}$ ratio for low-mass protostars ($2.2\pm1.2\times10^{-2}$) and low-mass prestellar cores ($3.4\pm1.9\times10^{-2}$).

\item For $\rm (D/H)_{CH_2DOH}$, significantly higher values than $\rm (D/H)_{CH_3OH}$ are found ranging from 0.1--1 with an average of $2.0\pm0.8\times10^{-1}$. The latter is good agreement with results on low-mass protostars and suggests that about 1/5 singly deuterated methanol molecules gets successively deuterated further independent of the mass of the system.

\item 
Based on a comparison with GRAINOBLE models in the literature, the lower $\rm (D/H)_{CH_3OH}$ ratios toward high-mass protostars suggest either a temperature of $\gtrsim 20$~K in the high-mass prestellar phases or a short lifetime ($\lesssim t_\mathrm{FF}$) of the high-mass prestellar phases. This is in strong contrast with the low-mass sources for which the higher $\rm (D/H)_{CH_3OH}$ ratio can only be achieved when the low-mass prestellar phases are both cold ($<15$~K) and long lived ($\geq t_\mathrm{FF}$). 
\end{itemize}

\noindent
This work demonstrates that the deuteration of the CH$_3$-group of methanol as measured toward protostellar systems could be used to probe the physical conditions (e.g., temperature) of the prestellar phases. 
The discrepancy in $\rm (D/H)_{CH_3OH}$ between low-mass and high-mass sources indicates that the physical conditions are already different before the onset of star formation. Additional observations of multiple deuterated methanol isotopologues (e.g., CHD$_2$OH, CD$_3$OH) as well as CH$_3$OD will shed further light on the efficiency of methanol deuteration between low-mass and high-mass systems. In combination with additional modeling studies \citep[such as those performed by][]{Bogelund2018,Taquet2019,Kulterer2022}, this can provide further insight on the relevant deuterium chemistry and how the D/H ratio varies across the protostellar mass range.

\begin{acknowledgements}
The authors would like to thank the anonymous referee for their constructive comments on the manuscript and L. Coudert for discussions on the CH$_2$DOH spectroscopy. This paper makes use of the following ALMA data: ADS/JAO.ALMA\#2017.1.01174.S, ADS/JAO.ALMA\#2019.1.00195.L. ALMA is a partnership of ESO (representing its member states), NSF (USA) and NINS (Japan), together with NRC (Canada), MOST and ASIAA (Taiwan), and KASI (Republic of Korea), in cooperation with the Republic of Chile. The Joint ALMA Observatory is operated by ESO, AUI/NRAO and NAOJ. 
Astrochemistry in Leiden is supported by the Netherlands Research School for Astronomy (NOVA), by funding from the European Research Council (ERC) under the European Union’s Horizon 2020 research and innovation programme (grant agreement No. 101019751 MOLDISK), and by the Dutch Research Council (NWO) grants TOP-1 614.001.751, 648.000.022, and 618.000.001. Support by the Danish National Research Foundation through the Center of Excellence “InterCat” (Grant agreement no.: DNRF150) is also acknowledged. 
\end{acknowledgements}


\bibliographystyle{aa}
\bibliography{refs}

\newcommand{\noop}[1]{}
\begin{thebibliography}{92}
\expandafter\ifx\csname natexlab\endcsname\relax\def\natexlab#1{#1}\fi

\bibitem[{{Aikawa} \& {Herbst}(1999)}]{Aikawa1999}
{Aikawa}, Y. \& {Herbst}, E. 1999, \apj, 526, 314

\bibitem[{{Ambrose} {et~al.}(2021){Ambrose}, {Shirley}, \&
  {Scibelli}}]{Ambrose2021}
{Ambrose}, H.~E., {Shirley}, Y.~L., \& {Scibelli}, S. 2021, \mnras, 501, 347

\bibitem[{{Belloche} {et~al.}(2016){Belloche}, {M{\"u}ller}, {Garrod}, \&
  {Menten}}]{Belloche2016}
{Belloche}, A., {M{\"u}ller}, H.~S.~P., {Garrod}, R.~T., \& {Menten}, K.~M.
  2016, \aap, 587, A91

\bibitem[{{Bianchi} {et~al.}(2020){Bianchi}, {Chandler}, {Ceccarelli},
  {Codella}, {Sakai}, {L{\'o}pez-Sepulcre}, {Maud}, {Moellenbrock}, {Svoboda},
  {Watanabe}, {Sakai}, {M{\'e}nard}, {Aikawa}, {Alves}, {Balucani}, {Bouvier},
  {Caselli}, {Caux}, {Charnley}, {Choudhury}, {De Simone}, {Dulieu},
  {Dur{\'a}n}, {Evans}, {Favre}, {Fedele}, {Feng}, {Fontani}, {Francis},
  {Hama}, {Hanawa}, {Herbst}, {Hirota}, {Imai}, {Isella}, {Jim{\'e}nez-Serra},
  {Johnstone}, {Kahane}, {Lefloch}, {Loinard}, {Maureira}, {Mercimek},
  {Miotello}, {Mori}, {Nakatani}, {Nomura}, {Oba}, {Ohashi}, {Okoda},
  {Ospina-Zamudio}, {Oya}, {Pineda}, {Podio}, {Rimola}, {Cox}, {Shirley},
  {Taquet}, {Testi}, {Vastel}, {Viti}, {Watanabe}, {Witzel}, {Xue}, {Zhang},
  {Zhao}, \& {Yamamoto}}]{Bianchi2020}
{Bianchi}, E., {Chandler}, C.~J., {Ceccarelli}, C., {et~al.} 2020, \mnras, 498,
  L87

\bibitem[{{Bianchi} {et~al.}(2017{\natexlab{a}}){Bianchi}, {Codella},
  {Ceccarelli}, {Fontani}, {Testi}, {Bachiller}, {Lefloch}, {Podio}, \&
  {Taquet}}]{Bianchi2017_SVS13A}
{Bianchi}, E., {Codella}, C., {Ceccarelli}, C., {et~al.} 2017{\natexlab{a}},
  \mnras, 467, 3011

\bibitem[{{Bianchi} {et~al.}(2017{\natexlab{b}}){Bianchi}, {Codella},
  {Ceccarelli}, {Taquet}, {Cabrit}, {Bacciotti}, {Bachiller}, {Chapillon},
  {Gueth}, {Gusdorf}, {Lefloch}, {Leurini}, {Podio}, {Rygl}, {Tabone}, \&
  {Tafalla}}]{Bianchi2017_HH212}
{Bianchi}, E., {Codella}, C., {Ceccarelli}, C., {et~al.} 2017{\natexlab{b}},
  \aap, 606, L7

\bibitem[{{Bizzocchi} {et~al.}(2014){Bizzocchi}, {Caselli}, {Spezzano}, \&
  {Leonardo}}]{Bizzocchi2014}
{Bizzocchi}, L., {Caselli}, P., {Spezzano}, S., \& {Leonardo}, E. 2014, \aap,
  569, A27

\bibitem[{{B{\o}gelund} {et~al.}(2019){B{\o}gelund}, {Barr}, {Taquet},
  {Ligterink}, {Persson}, {Hogerheijde}, \& {van Dishoeck}}]{Bogelund2019}
{B{\o}gelund}, E.~G., {Barr}, A.~G., {Taquet}, V., {et~al.} 2019, \aap, 628, A2

\bibitem[{{B{\o}gelund} {et~al.}(2018){B{\o}gelund}, {McGuire}, {Ligterink},
  {Taquet}, {Brogan}, {Hunter}, {Pearson}, {Hogerheijde}, \& {van
  Dishoeck}}]{Bogelund2018}
{B{\o}gelund}, E.~G., {McGuire}, B.~A., {Ligterink}, N. F.~W., {et~al.} 2018,
  \aap, 615, A88

\bibitem[{{Brown} \& {Millar}(1989)}]{Brown1989}
{Brown}, P.~D. \& {Millar}, T.~J. 1989, \mnras, 237, 661

\bibitem[{{Caselli} \& {Ceccarelli}(2012)}]{Caselli2012}
{Caselli}, P. \& {Ceccarelli}, C. 2012, \aapr, 20, 56

\bibitem[{{Ceccarelli} {et~al.}(2014){Ceccarelli}, {Caselli},
  {Bockel{\'e}e-Morvan}, {Mousis}, {Pizzarello}, {Robert}, \&
  {Semenov}}]{Ceccarelli2014}
{Ceccarelli}, C., {Caselli}, P., {Bockel{\'e}e-Morvan}, D., {et~al.} 2014, in
  Protostars and Planets VI, ed. H.~{Beuther}, R.~S. {Klessen}, C.~P.
  {Dullemond}, \& T.~{Henning}, 859

\bibitem[{{Chahine} {et~al.}(2022){Chahine}, {L{\'o}pez-Sepulcre}, {Neri},
  {Ceccarelli}, {Mercimek}, {Codella}, {Bouvier}, {Bianchi}, {Favre}, {Podio},
  {Alves}, {Sakai}, \& {Yamamoto}}]{Chahine2022}
{Chahine}, L., {L{\'o}pez-Sepulcre}, A., {Neri}, R., {et~al.} 2022, \aap, 657,
  A78

\bibitem[{{Coudert} {et~al.}(2021){Coudert}, {Motiyenko}, {Margul{\`e}s}, \&
  {Tchana Kwabia}}]{Coudert2021}
{Coudert}, L.~H., {Motiyenko}, R.~A., {Margul{\`e}s}, L., \& {Tchana Kwabia},
  F. 2021, Journal of Molecular Spectroscopy, 381, 111515

\bibitem[{{Coudert} {et~al.}(2014){Coudert}, {Zemouli}, {Motiyenko},
  {Margul{\`e}s}, \& {Klee}}]{Coudert2014}
{Coudert}, L.~H., {Zemouli}, M., {Motiyenko}, R.~A., {Margul{\`e}s}, L., \&
  {Klee}, S. 2014, \jcp, 140, 064307

\bibitem[{{Coutens} {et~al.}(2014){Coutens}, {J{\o}rgensen}, {Persson}, {van
  Dishoeck}, {Vastel}, \& {Taquet}}]{Coutens2014}
{Coutens}, A., {J{\o}rgensen}, J.~K., {Persson}, M.~V., {et~al.} 2014, \apjl,
  792, L5

\bibitem[{{Dartois} {et~al.}(2003){Dartois}, {Thi}, {Geballe}, {Deboffle},
  {d'Hendecourt}, \& {van Dishoeck}}]{Dartois2003}
{Dartois}, E., {Thi}, W.~F., {Geballe}, T.~R., {et~al.} 2003, \aap, 399, 1009

\bibitem[{{Drozdovskaya} {et~al.}(2022){Drozdovskaya}, {Coudert},
  {Margul{\`e}s}, {Coutens}, {J{\o}rgensen}, \& {Manigand}}]{Drozdovskaya2022}
{Drozdovskaya}, M.~N., {Coudert}, L.~H., {Margul{\`e}s}, L., {et~al.} 2022,
  \aap, 659, A69

\bibitem[{{Drozdovskaya} {et~al.}(2021){Drozdovskaya}, {Schroeder I}, {Rubin},
  {Altwegg}, {van Dishoeck}, {Kulterer}, {De Keyser}, {Fuselier}, \&
  {Combi}}]{Drozdovskaya2021}
{Drozdovskaya}, M.~N., {Schroeder I}, I. R.~H.~G., {Rubin}, M., {et~al.} 2021,
  \mnras, 500, 4901

\bibitem[{{Elia} {et~al.}(2021){Elia}, {Merello}, {Molinari}, {Schisano},
  {Zavagno}, {Russeil}, {M{\`e}ge}, {Martin}, {Olmi}, {Pestalozzi}, {Plume},
  {Ragan}, {Benedettini}, {Eden}, {Moore}, {Noriega-Crespo}, {Paladini},
  {Palmeirim}, {Pezzuto}, {Pilbratt}, {Rygl}, {Schilke}, {Strafella}, {Tan},
  {Traficante}, {Baldeschi}, {Bally}, {Giorgio}, {Fiorellino}, {Liu}, {Piazzo},
  \& {Polychroni}}]{Elia2021}
{Elia}, D., {Merello}, M., {Molinari}, S., {et~al.} 2021, \mnras, 504, 2742

\bibitem[{{Elia} {et~al.}(2017){Elia}, {Molinari}, {Schisano}, {Pestalozzi},
  {Pezzuto}, {Merello}, {Noriega-Crespo}, {Moore}, {Russeil}, {Mottram},
  {Paladini}, {Strafella}, {Benedettini}, {Bernard}, {Di Giorgio}, {Eden},
  {Fukui}, {Plume}, {Bally}, {Martin}, {Ragan}, {Jaffa}, {Motte}, {Olmi},
  {Schneider}, {Testi}, {Wyrowski}, {Zavagno}, {Calzoletti}, {Faustini},
  {Natoli}, {Palmeirim}, {Piacentini}, {Piazzo}, {Pilbratt}, {Polychroni},
  {Baldeschi}, {Beltr{\'a}n}, {Billot}, {Cambr{\'e}sy}, {Cesaroni},
  {Garc{\'\i}a-Lario}, {Hoare}, {Huang}, {Joncas}, {Liu}, {Maiolo}, {Marsh},
  {Maruccia}, {M{\`e}ge}, {Peretto}, {Rygl}, {Schilke}, {Thompson},
  {Traficante}, {Umana}, {Veneziani}, {Ward-Thompson}, {Whitworth}, {Arab},
  {Bandieramonte}, {Becciani}, {Brescia}, {Buemi}, {Bufano}, {Butora},
  {Cavuoti}, {Costa}, {Fiorellino}, {Hajnal}, {Hayakawa}, {Kacsuk}, {Leto}, {Li
  Causi}, {Marchili}, {Martinavarro-Armengol}, {Mercurio}, {Molinaro},
  {Riccio}, {Sano}, {Sciacca}, {Tachihara}, {Torii}, {Trigilio}, {Vitello}, \&
  {Yamamoto}}]{Elia2017}
{Elia}, D., {Molinari}, S., {Schisano}, E., {et~al.} 2017, \mnras, 471, 100

\bibitem[{{Endres} {et~al.}(2016){Endres}, {Schlemmer}, {Schilke}, {Stutzki},
  \& {M{\"u}ller}}]{Endres2016}
{Endres}, C.~P., {Schlemmer}, S., {Schilke}, P., {Stutzki}, J., \&
  {M{\"u}ller}, H. S.~P. 2016, Journal of Molecular Spectroscopy, 327, 95

\bibitem[{{Faure} {et~al.}(2015){Faure}, {Faure}, {Theul{\'e}}, {Quirico}, \&
  {Schmitt}}]{Faure2015}
{Faure}, A., {Faure}, M., {Theul{\'e}}, P., {Quirico}, E., \& {Schmitt}, B.
  2015, \aap, 584, A98

\bibitem[{{Fisher} {et~al.}(2007){Fisher}, {Paciga}, {Xu}, {Zhao}, {Moruzzi},
  \& {Lees}}]{Fisher2007}
{Fisher}, J., {Paciga}, G., {Xu}, L.-H., {et~al.} 2007, Journal of Molecular
  Spectroscopy, 245, 7

\bibitem[{{Fontani} {et~al.}(2015){Fontani}, {Busquet}, {Palau}, {Caselli},
  {S{\'a}nchez-Monge}, {Tan}, \& {Audard}}]{Fontani2015}
{Fontani}, F., {Busquet}, G., {Palau}, A., {et~al.} 2015, \aap, 575, A87

\bibitem[{{Fuchs} {et~al.}(2009){Fuchs}, {Cuppen}, {Ioppolo}, {Romanzin},
  {Bisschop}, {Andersson}, {van Dishoeck}, \& {Linnartz}}]{Fuchs2009}
{Fuchs}, G.~W., {Cuppen}, H.~M., {Ioppolo}, S., {et~al.} 2009, \aap, 505, 629

\bibitem[{{Fuente} {et~al.}(2014){Fuente}, {Cernicharo}, {Caselli}, {McCoey},
  {Johnstone}, {Fich}, {van Kempen}, {Palau}, {Y{\i}ld{\i}z}, {Tercero}, \&
  {L{\'o}pez}}]{Fuente2014}
{Fuente}, A., {Cernicharo}, J., {Caselli}, P., {et~al.} 2014, \aap, 568, A65

\bibitem[{{Furuya} {et~al.}(2016){Furuya}, {van Dishoeck}, \&
  {Aikawa}}]{Furuya2016}
{Furuya}, K., {van Dishoeck}, E.~F., \& {Aikawa}, Y. 2016, \aap, 586, A127

\bibitem[{{Hasegawa} \& {Herbst}(1993)}]{Hasegawa1993}
{Hasegawa}, T.~I. \& {Herbst}, E. 1993, \mnras, 263, 589

\bibitem[{{Hidaka} {et~al.}(2009){Hidaka}, {Watanabe}, {Kouchi}, \&
  {Watanabe}}]{Hidaka2009}
{Hidaka}, H., {Watanabe}, M., {Kouchi}, A., \& {Watanabe}, N. 2009, \apj, 702,
  291

\bibitem[{{Hsu} {et~al.}(2022){Hsu}, {Liu}, {Liu}, {Sahu}, {Lee}, {Tatematsu},
  {Kim}, {Hirano}, {Yang}, {Johnstone}, {Liu}, {Juvela}, {Bronfman}, {Chen},
  {Dutta}, {Eden}, {Jhan}, {Kuan}, {Lee}, {Lee}, {Li}, {Liu}, {Qin},
  {Sanhueza}, {Shang}, {Soam}, {Traficante}, \& {Zhou}}]{Hsu2022}
{Hsu}, S.-Y., {Liu}, S.-Y., {Liu}, T., {et~al.} 2022, \apj, 927, 218

\bibitem[{{Ilyushin} {et~al.}(2022){Ilyushin}, {M{\"u}ller}, {J{\o}rgensen},
  {Bauerecker}, {Maul}, {Bakhmat}, {Alekseev}, {Dorovskaya}, {Vlasenko},
  {Lewen}, {Schlemmer}, {Berezkin}, \& {Lees}}]{Ilyushin2022}
{Ilyushin}, V.~V., {M{\"u}ller}, H.~S.~P., {J{\o}rgensen}, J.~K., {et~al.}
  2022, \aap, 658, A127

\bibitem[{{Jacobsen} {et~al.}(2019){Jacobsen}, {J{\o}rgensen}, {Di Francesco},
  {Evans}, {Choi}, \& {Lee}}]{Jacobsen2019}
{Jacobsen}, S.~K., {J{\o}rgensen}, J.~K., {Di Francesco}, J., {et~al.} 2019,
  \aap, 629, A29

\bibitem[{{Jensen} {et~al.}(2021){Jensen}, {J{\o}rgensen}, {Kristensen},
  {Coutens}, {van Dishoeck}, {Furuya}, {Harsono}, \& {Persson}}]{Jensen2021}
{Jensen}, S.~S., {J{\o}rgensen}, J.~K., {Kristensen}, L.~E., {et~al.} 2021,
  \aap, 650, A172

\bibitem[{{Jensen} {et~al.}(2019){Jensen}, {J{\o}rgensen}, {Kristensen},
  {Furuya}, {Coutens}, {van Dishoeck}, {Harsono}, \& {Persson}}]{Jensen2019}
{Jensen}, S.~S., {J{\o}rgensen}, J.~K., {Kristensen}, L.~E., {et~al.} 2019,
  \aap, 631, A25

\bibitem[{{J{\o}rgensen} {et~al.}(2018){J{\o}rgensen}, {M{\"u}ller}, {Calcutt},
  {Coutens}, {Drozdovskaya}, {{\"O}berg}, {Persson}, {Taquet}, {van Dishoeck},
  \& {Wampfler}}]{Jorgensen2018}
{J{\o}rgensen}, J.~K., {M{\"u}ller}, H.~S.~P., {Calcutt}, H., {et~al.} 2018,
  \aap, 620, A170

\bibitem[{{Kulterer} {et~al.}(2022){Kulterer}, {Drozdovskaya}, {Antonellini},
  {Walsh}, \& {Millar}}]{Kulterer2022}
{Kulterer}, B.~M., {Drozdovskaya}, M.~N., {Antonellini}, S., {Walsh}, C., \&
  {Millar}, T.~J. 2022, ACS Earth and Space Chemistry, 6, 1171

\bibitem[{{Lattanzi} {et~al.}(2020){Lattanzi}, {Bizzocchi}, {Vasyunin},
  {Harju}, {Giuliano}, {Vastel}, \& {Caselli}}]{Lattanzi2020}
{Lattanzi}, V., {Bizzocchi}, L., {Vasyunin}, A.~I., {et~al.} 2020, \aap, 633,
  A118

\bibitem[{{Lee} {et~al.}(2019{\natexlab{a}}){Lee}, {Codella}, {Li}, \&
  {Liu}}]{Lee2019_HH212}
{Lee}, C.-F., {Codella}, C., {Li}, Z.-Y., \& {Liu}, S.-Y. 2019{\natexlab{a}},
  \apj, 876, 63

\bibitem[{{Lee} {et~al.}(2019{\natexlab{b}}){Lee}, {Lee}, {Baek}, {Aikawa},
  {Cieza}, {Yoon}, {Herczeg}, {Johnstone}, \& {Casassus}}]{Lee2019_V883}
{Lee}, J.-E., {Lee}, S., {Baek}, G., {et~al.} 2019{\natexlab{b}}, Nature
  Astronomy, 3, 314

\bibitem[{{Ligterink} {et~al.}(2021){Ligterink}, {Ahmadi}, {Coutens},
  {Tychoniec}, {Calcutt}, {van Dishoeck}, {Linnartz}, {J{\o}rgensen}, {Garrod},
  \& {Bouwman}}]{Ligterink2021}
{Ligterink}, N.~F.~W., {Ahmadi}, A., {Coutens}, A., {et~al.} 2021, \aap, 647,
  A87

\bibitem[{{Ligterink} {et~al.}(2022){Ligterink}, {Ahmadi}, {Luitel}, {Coutens},
  {Calcutt}, {Tychoniec}, {Linnartz}, {J{\o}rgensen}, {Garrod}, \&
  {Bouwman}}]{Ligterink2022}
{Ligterink}, N. F.~W., {Ahmadi}, A., {Luitel}, B., {et~al.} 2022, ACS Earth and
  Space Chemistry, 6, 455

\bibitem[{{Linsky} {et~al.}(2006){Linsky}, {Draine}, {Moos}, {Jenkins}, {Wood},
  {Oliveira}, {Blair}, {Friedman}, {Gry}, {Knauth}, {Kruk}, {Lacour}, {Lehner},
  {Redfield}, {Shull}, {Sonneborn}, \& {Williger}}]{Linsky2006}
{Linsky}, J.~L., {Draine}, B.~T., {Moos}, H.~W., {et~al.} 2006, \apj, 647, 1106

\bibitem[{{Manigand} {et~al.}(2020){Manigand}, {J{\o}rgensen}, {Calcutt},
  {M{\"u}ller}, {Ligterink}, {Coutens}, {Drozdovskaya}, {van Dishoeck}, \&
  {Wampfler}}]{Manigand2020}
{Manigand}, S., {J{\o}rgensen}, J.~K., {Calcutt}, H., {et~al.} 2020, \aap, 635,
  A48

\bibitem[{{Mart{\'\i}n-Dom{\'e}nech} {et~al.}(2021){Mart{\'\i}n-Dom{\'e}nech},
  {Bergner}, {{\"O}berg}, {Carpenter}, {Law}, {Huang}, {J{\o}rgensen},
  {Schwarz}, \& {Wilner}}]{Martin-domenech2021}
{Mart{\'\i}n-Dom{\'e}nech}, R., {Bergner}, J.~B., {{\"O}berg}, K.~I., {et~al.}
  2021, \apj, 923, 155

\bibitem[{{Mart{\'\i}n-Dom{\'e}nech} {et~al.}(2019){Mart{\'\i}n-Dom{\'e}nech},
  {Bergner}, {{\"O}berg}, \& {J{\o}rgensen}}]{Martin-domenech2019}
{Mart{\'\i}n-Dom{\'e}nech}, R., {Bergner}, J.~B., {{\"O}berg}, K.~I., \&
  {J{\o}rgensen}, J.~K. 2019, \apj, 880, 130

\bibitem[{{McMullin} {et~al.}(2007){McMullin}, {Waters}, {Schiebel}, {Young},
  \& {Golap}}]{McMullin2007}
{McMullin}, J.~P., {Waters}, B., {Schiebel}, D., {Young}, W., \& {Golap}, K.
  2007, in Astronomical Society of the Pacific Conference Series, Vol. 376,
  Astronomical Data Analysis Software and Systems XVI, ed. R.~A. {Shaw},
  F.~{Hill}, \& D.~J. {Bell}, 127

\bibitem[{{M{\`e}ge} {et~al.}(2021){M{\`e}ge}, {Russeil}, {Zavagno}, {Elia},
  {Molinari}, {Brunt}, {Butora}, {Cambresy}, {Di Giorgio}, {Fenouillet},
  {Fukui}, {Lambert}, {Makai}, {Merello}, {Meunier}, {Molinaro}, {Moreau},
  {Pezzuto}, {Poulin}, {Schisano}, \& {Schuller}}]{Mege2021}
{M{\`e}ge}, P., {Russeil}, D., {Zavagno}, A., {et~al.} 2021, \aap, 646, A74

\bibitem[{{Milam} {et~al.}(2005){Milam}, {Savage}, {Brewster}, {Ziurys}, \&
  {Wyckoff}}]{Milam2005}
{Milam}, S.~N., {Savage}, C., {Brewster}, M.~A., {Ziurys}, L.~M., \& {Wyckoff},
  S. 2005, \apj, 634, 1126

\bibitem[{{Molinari} {et~al.}(2010){Molinari}, {Swinyard}, {Bally}, {Barlow},
  {Bernard}, {Martin}, {Moore}, {Noriega-Crespo}, {Plume}, {Testi}, {Zavagno},
  {Abergel}, {Ali}, {Anderson}, {Andr{\'e}}, {Baluteau}, {Battersby},
  {Beltr{\'a}n}, {Benedettini}, {Billot}, {Blommaert}, {Bontemps}, {Boulanger},
  {Brand}, {Brunt}, {Burton}, {Calzoletti}, {Carey}, {Caselli}, {Cesaroni},
  {Cernicharo}, {Chakrabarti}, {Chrysostomou}, {Cohen}, {Compiegne}, {de
  Bernardis}, {de Gasperis}, {di Giorgio}, {Elia}, {Faustini}, {Flagey},
  {Fukui}, {Fuller}, {Ganga}, {Garcia-Lario}, {Glenn}, {Goldsmith}, {Griffin},
  {Hoare}, {Huang}, {Ikhenaode}, {Joblin}, {Joncas}, {Juvela}, {Kirk},
  {Lagache}, {Li}, {Lim}, {Lord}, {Marengo}, {Marshall}, {Masi}, {Massi},
  {Matsuura}, {Minier}, {Miville-Desch{\^e}nes}, {Montier}, {Morgan}, {Motte},
  {Mottram}, {M{\"u}ller}, {Natoli}, {Neves}, {Olmi}, {Paladini}, {Paradis},
  {Parsons}, {Peretto}, {Pestalozzi}, {Pezzuto}, {Piacentini}, {Piazzo},
  {Polychroni}, {Pomar{\`e}s}, {Popescu}, {Reach}, {Ristorcelli}, {Robitaille},
  {Robitaille}, {Rod{\'o}n}, {Roy}, {Royer}, {Russeil}, {Saraceno}, {Sauvage},
  {Schilke}, {Schisano}, {Schneider}, {Schuller}, {Schulz}, {Sibthorpe},
  {Smith}, {Smith}, {Spinoglio}, {Stamatellos}, {Strafella}, {Stringfellow},
  {Sturm}, {Taylor}, {Thompson}, {Traficante}, {Tuffs}, {Umana}, {Valenziano},
  {Vavrek}, {Veneziani}, {Viti}, {Waelkens}, {Ward-Thompson}, {White},
  {Wilcock}, {Wyrowski}, {Yorke}, \& {Zhang}}]{Molinari2010}
{Molinari}, S., {Swinyard}, B., {Bally}, J., {et~al.} 2010, \aap, 518, L100

\bibitem[{{M{\"u}ller} {et~al.}(2016){M{\"u}ller}, {Belloche}, {Xu}, {Lees},
  {Garrod}, {Walters}, {van Wijngaarden}, {Lewen}, {Schlemmer}, \&
  {Menten}}]{Muller2016}
{M{\"u}ller}, H. S.~P., {Belloche}, A., {Xu}, L.-H., {et~al.} 2016, \aap, 587,
  A92

\bibitem[{{M{\"u}ller} {et~al.}(2005){M{\"u}ller}, {Schl{\"o}der}, {Stutzki},
  \& {Winnewisser}}]{Muller2005}
{M{\"u}ller}, H. S.~P., {Schl{\"o}der}, F., {Stutzki}, J., \& {Winnewisser}, G.
  2005, Journal of Molecular Structure, 742, 215

\bibitem[{{M{\"u}ller} {et~al.}(2001){M{\"u}ller}, {Thorwirth}, {Roth}, \&
  {Winnewisser}}]{Muller2001}
{M{\"u}ller}, H.~S.~P., {Thorwirth}, S., {Roth}, D.~A., \& {Winnewisser}, G.
  2001, \aap, 370, L49

\bibitem[{{Nagaoka} {et~al.}(2005){Nagaoka}, {Watanabe}, \&
  {Kouchi}}]{Nagaoka2005}
{Nagaoka}, A., {Watanabe}, N., \& {Kouchi}, A. 2005, \apjl, 624, L29

\bibitem[{{Nagaoka} {et~al.}(2007){Nagaoka}, {Watanabe}, \&
  {Kouchi}}]{Nagaoka2007}
{Nagaoka}, A., {Watanabe}, N., \& {Kouchi}, A. 2007, Journal of Physical
  Chemistry A, 111, 3016

\bibitem[{{Nazari} {et~al.}(2022){Nazari}, {Meijerhof}, {van Gelder}, \& {van
  Dishoeck}}]{Nazari2022_ALMAGAL}
{Nazari}, P., {Meijerhof}, J., {van Gelder}, M., \& {van Dishoeck}, E.~F. 2022,
  \aap, submitted

\bibitem[{{Nazari} {et~al.}(2021){Nazari}, {van Gelder}, {van Dishoeck},
  {Tabone}, {van't Hoff}, {Ligterink}, {Beuther}, {Boogert}, {Caratti o
  Garatti}, {Klaassen}, {Linnartz}, {Taquet}, \& {Tychoniec}}]{Nazari2021}
{Nazari}, P., {van Gelder}, M.~L., {van Dishoeck}, E.~F., {et~al.} 2021, \aap,
  650, A150

\bibitem[{{Neill} {et~al.}(2013){Neill}, {Crockett}, {Bergin}, {Pearson}, \&
  {Xu}}]{Neill2013}
{Neill}, J.~L., {Crockett}, N.~R., {Bergin}, E.~A., {Pearson}, J.~C., \& {Xu},
  L.-H. 2013, \apj, 777, 85

\bibitem[{{Noble} {et~al.}(2012){Noble}, {Congiu}, {Dulieu}, \&
  {Fraser}}]{Noble2012}
{Noble}, J.~A., {Congiu}, E., {Dulieu}, F., \& {Fraser}, H.~J. 2012, \mnras,
  421, 768

\bibitem[{{Ohno} {et~al.}(2022){Ohno}, {Oyama}, {Tamanai}, {Zeng}, {Watanabe},
  {Nakatani}, {Sakai}, \& {Sakai}}]{Ohno2022}
{Ohno}, Y., {Oyama}, T., {Tamanai}, A., {et~al.} 2022, \apj, 932, 101

\bibitem[{{Ospina-Zamudio} {et~al.}(2018){Ospina-Zamudio}, {Lefloch},
  {Ceccarelli}, {Kahane}, {Favre}, {L{\'o}pez-Sepulcre}, \&
  {Montarges}}]{Ospina-Zamudio2018}
{Ospina-Zamudio}, J., {Lefloch}, B., {Ceccarelli}, C., {et~al.} 2018, \aap,
  618, A145

\bibitem[{{Parise} {et~al.}(2004){Parise}, {Castets}, {Herbst}, {Caux},
  {Ceccarelli}, {Mukhopadhyay}, \& {Tielens}}]{Parise2004}
{Parise}, B., {Castets}, A., {Herbst}, E., {et~al.} 2004, \aap, 416, 159

\bibitem[{{Parise} {et~al.}(2002){Parise}, {Ceccarelli}, {Tielens}, {Herbst},
  {Lefloch}, {Caux}, {Castets}, {Mukhopadhyay}, {Pagani}, \&
  {Loinard}}]{Parise2002}
{Parise}, B., {Ceccarelli}, C., {Tielens}, A.~G.~G.~M., {et~al.} 2002, \aap,
  393, L49

\bibitem[{{Pearson} {et~al.}(2012){Pearson}, {Yu}, \& {Drouin}}]{Pearson2012}
{Pearson}, J.~C., {Yu}, S., \& {Drouin}, B.~J. 2012, Journal of Molecular
  Spectroscopy, 280, 119

\bibitem[{{Perotti} {et~al.}(2021){Perotti}, {J{\o}rgensen}, {Fraser},
  {Suutarinen}, {Kristensen}, {Rocha}, {Bjerkeli}, \&
  {Pontoppidan}}]{Perotti2021}
{Perotti}, G., {J{\o}rgensen}, J.~K., {Fraser}, H.~J., {et~al.} 2021, \aap,
  650, A168

\bibitem[{{Perotti} {et~al.}(2020){Perotti}, {Rocha}, {J{\o}rgensen},
  {Kristensen}, {Fraser}, \& {Pontoppidan}}]{Perotti2020}
{Perotti}, G., {Rocha}, W.~R.~M., {J{\o}rgensen}, J.~K., {et~al.} 2020, \aap,
  643, A48

\bibitem[{{Persson} {et~al.}(2018){Persson}, {J{\o}rgensen}, {M{\"u}ller},
  {Coutens}, {van Dishoeck}, {Taquet}, {Calcutt}, {van der Wiel}, {Bourke}, \&
  {Wampfler}}]{Persson2018}
{Persson}, M.~V., {J{\o}rgensen}, J.~K., {M{\"u}ller}, H.~S.~P., {et~al.} 2018,
  \aap, 610, A54

\bibitem[{{Persson} {et~al.}(2014){Persson}, {J{\o}rgensen}, {van Dishoeck}, \&
  {Harsono}}]{Persson2014}
{Persson}, M.~V., {J{\o}rgensen}, J.~K., {van Dishoeck}, E.~F., \& {Harsono},
  D. 2014, \aap, 563, A74

\bibitem[{{Pickett} {et~al.}(1998){Pickett}, {Poynter}, {Cohen}, {Delitsky},
  {Pearson}, \& {M{\"u}ller}}]{Pickett1998}
{Pickett}, H.~M., {Poynter}, R.~L., {Cohen}, E.~A., {et~al.} 1998, \jqsrt, 60,
  883

\bibitem[{{Prodanovi{\'c}} {et~al.}(2010){Prodanovi{\'c}}, {Steigman}, \&
  {Fields}}]{Prodanovic2010}
{Prodanovi{\'c}}, T., {Steigman}, G., \& {Fields}, B.~D. 2010, \mnras, 406,
  1108

\bibitem[{{Ratajczak} {et~al.}(2009){Ratajczak}, {Quirico}, {Faure}, {Schmitt},
  \& {Ceccarelli}}]{Ratajczak2009}
{Ratajczak}, A., {Quirico}, E., {Faure}, A., {Schmitt}, B., \& {Ceccarelli}, C.
  2009, \aap, 496, L21

\bibitem[{{Roberts} {et~al.}(2003){Roberts}, {Herbst}, \&
  {Millar}}]{Roberts2003}
{Roberts}, H., {Herbst}, E., \& {Millar}, T.~J. 2003, \apjl, 591, L41

\bibitem[{{Santos} {et~al.}(2022){Santos}, {Chuang}, {Lamberts}, {Fedoseev},
  {Ioppolo}, \& {Linnartz}}]{Santos2022}
{Santos}, J.~C., {Chuang}, K.-J., {Lamberts}, T., {et~al.} 2022, \apjl, 931,
  L33

\bibitem[{{Simons} {et~al.}(2020){Simons}, {Lamberts}, \&
  {Cuppen}}]{Simons2020}
{Simons}, M.~A.~J., {Lamberts}, T., \& {Cuppen}, H.~M. 2020, \aap, 634, A52

\bibitem[{{Taquet} {et~al.}(2019){Taquet}, {Bianchi}, {Codella}, {Persson},
  {Ceccarelli}, {Cabrit}, {J{\o}rgensen}, {Kahane}, {L{\'o}pez-Sepulcre}, \&
  {Neri}}]{Taquet2019}
{Taquet}, V., {Bianchi}, E., {Codella}, C., {et~al.} 2019, \aap, 632, A19

\bibitem[{{Taquet} {et~al.}(2012){Taquet}, {Ceccarelli}, \&
  {Kahane}}]{Taquet2012}
{Taquet}, V., {Ceccarelli}, C., \& {Kahane}, C. 2012, \aap, 538, A42

\bibitem[{{Taquet} {et~al.}(2014){Taquet}, {Charnley}, \&
  {Sipil{\"a}}}]{Taquet2014}
{Taquet}, V., {Charnley}, S.~B., \& {Sipil{\"a}}, O. 2014, \apj, 791, 1

\bibitem[{{Taquet} {et~al.}(2013){Taquet}, {Peters}, {Kahane}, {Ceccarelli},
  {L{\'o}pez-Sepulcre}, {Toubin}, {Duflot}, \& {Wiesenfeld}}]{Taquet2013}
{Taquet}, V., {Peters}, P.~S., {Kahane}, C., {et~al.} 2013, \aap, 550, A127

\bibitem[{{Tielens}(1983)}]{Tielens1983}
{Tielens}, A.~G.~G.~M. 1983, \aap, 119, 177

\bibitem[{{Tielens}(2013)}]{Tielens2013}
{Tielens}, A.~G.~G.~M. 2013, Reviews of Modern Physics, 85, 1021

\bibitem[{{van der Walt} {et~al.}(2021){van der Walt}, {Kristensen},
  {J{\o}rgensen}, {Calcutt}, {Manigand}, {el Akel}, {Garrod}, \&
  {Qiu}}]{vanderWalt2021}
{van der Walt}, S.~J., {Kristensen}, L.~E., {J{\o}rgensen}, J.~K., {et~al.}
  2021, \aap, 655, A86

\bibitem[{{van Dishoeck} {et~al.}(1995){van Dishoeck}, {Blake}, {Jansen}, \&
  {Groesbeck}}]{vanDishoeck1995}
{van Dishoeck}, E.~F., {Blake}, G.~A., {Jansen}, D.~J., \& {Groesbeck}, T.~D.
  1995, \apj, 447, 760

\bibitem[{{van Gelder} {et~al.}(2022){van Gelder}, {Nazari}, {Tabone},
  {Ahmadi}, {van Dishoeck}, {Beltr{\'a}n}, {Fuller}, {Sakai},
  {S{\'a}nchez-Monge}, {Schilke}, {Yang}, \& {Zhang}}]{vanGelder2022}
{van Gelder}, M.~L., {Nazari}, P., {Tabone}, B., {et~al.} 2022, \aap, 662, A67

\bibitem[{{van Gelder} {et~al.}(2020){van Gelder}, {Tabone}, {Tychoniec}, {van
  Dishoeck}, {Beuther}, {Boogert}, {Caratti o Garatti}, {Klaassen}, {Linnartz},
  {M{\"u}ller}, \& {Taquet}}]{vanGelder2020}
{van Gelder}, M.~L., {Tabone}, B., {Tychoniec}, {\L}., {et~al.} 2020, \aap,
  639, A87

\bibitem[{{van't Hoff} {et~al.}(2022){van't Hoff}, {Harsono}, {van Gelder},
  {Hsieh}, {Tobin}, {Jensen}, {Hirano}, {J{\o}rgensen}, {Bergin}, \& {van
  Dishoeck}}]{vantHoff2022}
{van't Hoff}, M. L.~R., {Harsono}, D., {van Gelder}, M.~L., {et~al.} 2022,
  \apj, 924, 5

\bibitem[{{Vastel} {et~al.}(2015){Vastel}, {Bottinelli}, {Caux}, {Glorian}, \&
  {Boiziot}}]{Vastel2015}
{Vastel}, C., {Bottinelli}, S., {Caux}, E., {Glorian}, J.~M., \& {Boiziot}, M.
  2015, in SF2A-2015: Proceedings of the Annual meeting of the French Society
  of Astronomy and Astrophysics, 313--316

\bibitem[{{Watanabe} \& {Kouchi}(2002)}]{Watanabe2002}
{Watanabe}, N. \& {Kouchi}, A. 2002, \apjl, 571, L173

\bibitem[{{Watson}(1974)}]{Watson1974}
{Watson}, W.~D. 1974, \apj, 188, 35

\bibitem[{{Wilson} \& {Rood}(1994)}]{Wilson1994}
{Wilson}, T.~L. \& {Rood}, R. 1994, \araa, 32, 191

\bibitem[{{Xu} {et~al.}(2008){Xu}, {Fisher}, {Lees}, {Shi}, {Hougen},
  {Pearson}, {Drouin}, {Blake}, \& {Braakman}}]{Xu2008}
{Xu}, L.-H., {Fisher}, J., {Lees}, R.~M., {et~al.} 2008, Journal of Molecular
  Spectroscopy, 251, 305

\bibitem[{{Xu} \& {Lovas}(1997)}]{Xu1997}
{Xu}, L.-H. \& {Lovas}, F.~J. 1997, Journal of Physical and Chemical Reference
  Data, 26, 17

\bibitem[{{Yang} {et~al.}(2020){Yang}, {Evans}, {Smith}, {Lee}, {Tobin},
  {Terebey}, {Calcutt}, {J{\o}rgensen}, {Green}, \& {Bourke}}]{Yang2020}
{Yang}, Y.-L., {Evans}, Neal~J., I., {Smith}, A., {et~al.} 2020, \apj, 891, 61

\end{thebibliography}

\onecolumn
\begin{appendix}

\section{Transitions of CH$_3$OH and isotopologues}
\label{app:CH3OH_transitions}
\begin{longtable}{lrclccl}
\caption{Transitions of CH$_3$OH and isotopologues with $A_\mathrm{ij} > 10^{-6}$ and $E_\mathrm{up} < 1000$~K covered in the ALMAGAL (2019.1.00195.L) program. \label{tab:CH3OH_transitions}} \\
\hline\hline
Species & \multicolumn{3}{c}{Transition} & Frequency & $A_\mathrm{ij}$ & $E_\mathrm{up}$ \\
 & (J K L M & - & J K L M) & (GHz) & (s$^{-1}$) & (K) \\
\hline
\endfirsthead
\caption{continued.}\\
\hline\hline
Species & \multicolumn{3}{c}{Transition} & Frequency & $A_\mathrm{ij}$ & $E_\mathrm{up}$ \\
 & (J K L M & - & J K L M) & (GHz) & (s$^{-1}$) & (K) \\
\hline
\endhead
\hline
\endfoot
CH$_3$OH & 6 1 5 3 & - & 7 2 5 3 & 217.2992 & $4.3 \times 10^{-5}$ & 373.9 \\
 & 15 6 9 3 & - & 16 5 11 3 & 217.6427 & $1.9 \times 10^{-5}$ & 745.6 \\
 & 15 6 10 3 & - & 16 5 12 3 & 217.6427 & $1.9 \times 10^{-5}$ & 745.6 \\
 & 20 1 19 1 & - & 20 0 20 1 & 217.8865 & $3.4 \times 10^{-5}$ & 508.4 \\
 & 4 2 3 1 & - & 3 1 2 1 & 218.4401 & $4.7 \times 10^{-5}$ & 45.5 \\
 & 25 3 23 1 & - & 24 4 20 1 & 219.9837 & $2.0 \times 10^{-5}$ & 802.2 \\
 & 23 5 18 1 & - & 22 6 17 1 & 219.9937 & $1.7 \times 10^{-5}$ & 775.9 \\
 & 8 0 8 1 & - & 7 1 6 1 & 220.0786 & $2.5 \times 10^{-5}$ & 96.6 \\
 & 10 5 6 2 & - & 11 4 8 2 & 220.4013 & $1.1 \times 10^{-5}$ & 251.6 \\
\hline
$^{13}$CH$_3$OH & 14 1 13 -0 & - & 13 2 12 -0 & 217.0446 & $2.4 \times 10^{-5}$ & 254.3 \\
 & 10 2 8 +0 & - & 9 3 7 +0 & 217.3995 & $1.5 \times 10^{-5}$ & 162.4 \\
 & 17 7 11 +0 & - & 18 6 12 +0 & 220.3218 & $1.3 \times 10^{-5}$ & 592.3 \\
 & 17 7 10 -0 & - & 18 6 13 -0 & 220.3218 & $1.3 \times 10^{-5}$ & 592.3 \\
\hline
CH$_3^{18}$OH & 14 1 14 1 & - & 13 2 12 1 & 217.1729 & $1.7 \times 10^{-5}$ & 238.9 \\
 & 18 6 13 4 & - & 17 7 11 4 & 217.9223 & $1.5 \times 10^{-5}$ & 874.1 \\
 & 17 5 13 4 & - & 18 6 13 4 & 218.5521 & $3.2 \times 10^{-5}$ & 884.6 \\
 & 4 2 2 2 & - & 3 1 2 2 & 219.4078 & $4.6 \times 10^{-5}$ & 44.6 \\
 & 8 7 1 5 & - & 7 6 1 5 & 219.8433 & $2.8 \times 10^{-5}$ & 663.2 \\
 & 18 3 16 5 & - & 19 4 16 5 & 219.9572 & $5.1 \times 10^{-5}$ & 795.8 \\
 & 8 1 8 1 & - & 7 0 7 1 & 220.1951 & $3.6 \times 10^{-5}$ & 85.7 \\
\hline
CH$_2$DOH & 26 4 22 0 & - & 26 3 24 2 & 217.2664 & $2.0 \times 10^{-5}$ & 817.1 \\
 & 26 1 25 2 & - & 26 1 26 2 & 217.3300 & $1.1 \times 10^{-5}$ & 777.9 \\
 & 17 4 13 2 & - & 16 5 11 1 & 217.3436 & $5.2 \times 10^{-6}$ & 409.7 \\
 & 17 4 14 2 & - & 16 5 12 1 & 217.3593 & $5.3 \times 10^{-6}$ & 409.7 \\
 & 23 6 18 1 & - & 22 7 15 0 & 217.3818 & $6.4 \times 10^{-6}$ & 742.6 \\
 & 23 6 17 1 & - & 22 7 16 0 & 217.3825 & $6.4 \times 10^{-6}$ & 742.6 \\
 & 18 1 17 2 & - & 18 2 17 0 & 217.4479 & $1.8 \times 10^{-5}$ & 391.5 \\
 & 25 1 25 2 & - & 25 0 25 1 & 217.6429 & $4.7 \times 10^{-5}$ & 712.4 \\
 & 12 7 6 0 & - & 13 6 7 1 & 217.6446 & $2.5 \times 10^{-6}$ & 357.2 \\
 & 12 7 5 0 & - & 13 6 8 1 & 217.6446 & $2.5 \times 10^{-6}$ & 357.2 \\
 & 18 0 18 0 & - & 17 1 16 2 & 218.1095 & $8.9 \times 10^{-6}$ & 363.2 \\
 & 5 2 4 1 & - & 5 1 5 1 & 218.3164 & $9.1 \times 10^{-6}$ & 58.7 \\
 & 24 3 21 1 & - & 24 2 23 2 & 218.5348 & $3.9 \times 10^{-5}$ & 687.7 \\
 & 20 5 16 1 & - & 19 6 13 0 & 219.2043 & $1.5 \times 10^{-5}$ & 557.6 \\
 & 20 5 15 1 & - & 19 6 14 0 & 219.2061 & $1.5 \times 10^{-5}$ & 557.6 \\
 & 5 1 5 1 & - & 4 1 4 1 & 219.5515 & $7.0 \times 10^{-6}$ & 48.2 \\
 & 5 1 5 0 & - & 4 1 4 0 & 220.0718 & $3.3 \times 10^{-5}$ & 35.8 \\
 & 29 4 26 0 & - & 29 3 26 2 & 220.3492 & $2.5 \times 10^{-5}$ & 997.1 \\
 & 17 1 16 0 & - & 17 0 17 0 & 220.5526 & $3.8 \times 10^{-5}$ & 335.9 \\
 & 21 1 20 1 & - & 21 1 21 1 & 220.6256 & $2.0 \times 10^{-6}$ & 515.1 \\
 & 21 2 19 2 & - & 21 1 20 2 & 220.7358 & $3.9 \times 10^{-5}$ & 531.1 \\
\hline
CHD$_2$OH & 6 2 2 1 & - & 5 1 2 2 & 217.0702 & $1.8 \times 10^{-6}$ & 61.9 \\
 & 13 2 1 2 & - & 12 3 1 2 & 217.1181 & $8.8 \times 10^{-6}$ & 213.0 \\
 & 16 5 1 2 & - & 17 4 1 0 & 217.2651 & $2.4 \times 10^{-6}$ & 366.8 \\
 & 7 0 1 2 & - & 6 1 1 2 & 217.4912 & $2.9 \times 10^{-5}$ & 74.3 \\
 & 7 4 2 1 & - & 8 3 2 0 & 217.4946 & $4.4 \times 10^{-6}$ & 111.1 \\
 & 7 4 1 1 & - & 8 3 1 0 & 217.5430 & $4.4 \times 10^{-6}$ & 111.1 \\
 & 24 9 1 1 & - & 25 8 1 0 & 217.8034 & $9.0 \times 10^{-6}$ & 854.9 \\
 & 24 9 2 1 & - & 25 8 2 0 & 217.8034 & $9.0 \times 10^{-6}$ & 854.9 \\
 & 2 2 1 1 & - & 3 1 1 0 & 218.0092 & $5.5 \times 10^{-6}$ & 25.8 \\
 & 7 2 1 0 & - & 7 1 2 0 & 218.1279 & $1.2 \times 10^{-5}$ & 68.8 \\
 & 16 3 2 2 & - & 15 4 2 2 & 218.2323 & $6.9 \times 10^{-6}$ & 318.2 \\
 & 20 2 2 2 & - & 20 2 1 1 & 218.4156 & $9.0 \times 10^{-6}$ & 449.9 \\
 & 11 2 1 2 & - & 11 1 2 2 & 218.4824 & $2.3 \times 10^{-5}$ & 163.0 \\
 & 12 2 1 2 & - & 12 1 2 2 & 219.2181 & $4.9 \times 10^{-6}$ & 187.0 \\
 & 23 10 2 2 & - & 24 9 2 2 & 219.3323 & $5.1 \times 10^{-6}$ & 872.6 \\
 & 23 10 1 2 & - & 24 9 1 2 & 219.3323 & $5.1 \times 10^{-6}$ & 872.6 \\
 & 16 3 1 0 & - & 15 4 1 0 & 219.3377 & $6.5 \times 10^{-6}$ & 300.9 \\
 & 22 5 2 0 & - & 22 4 1 2 & 219.6514 & $2.6 \times 10^{-6}$ & 583.5 \\
 & 22 5 1 0 & - & 22 4 2 2 & 219.7983 & $2.5 \times 10^{-6}$ & 583.4 \\
 & 13 2 1 0 & - & 12 3 1 0 & 220.2430 & $4.7 \times 10^{-6}$ & 195.3 \\
 & 11 6 1 1 & - & 12 5 1 1 & 220.5567 & $1.1 \times 10^{-6}$ & 249.5 \\
 & 11 6 2 1 & - & 12 5 2 1 & 220.5569 & $1.1 \times 10^{-6}$ & 249.5 \\
\hline
\end{longtable}
\tablefoot{
The typical beam size is $\theta_\mathrm{beam}\sim1^{\prime\prime}$ and the typical rms is $\mathrm{rms_{line}}\sim0.2$~K.
}

\section{Observational details}
\label{app:Observational_details}

\renewcommand*{\arraystretch}{1.1}
\begin{landscape}
\footnotesize{
\begin{longtable}[c]{llllllllllll}
\caption{Column densities of $^{13}$CH$_3$OH, CH$_3^{18}$OH, CH$_3$OH, CH$_2$DOH, and CHD$_2$OH and derived methanol D/H ratios.\label{tab:D_H_columns}}\\
\hline\hline
Source & RA (J2000) & Dec (J2000) & $\theta_\mathrm{beam}$ & $T_\mathrm{ex}$ & $N_\mathrm{^{13}CH_3OH}$ & $N_\mathrm{CH_3^{18}OH}$ & $N_\mathrm{CH_3OH}$ & $N_\mathrm{CH_2DOH}$ & $N_\mathrm{CHD_2OH}$ & $\rm (D/H)_{CH_3OH}$\tablefootmark{1} & $\rm (D/H)_{CH_2DOH}$\tablefootmark{2} \\
& & & $^{\prime\prime}$ & K & cm$^{-2}$ & cm$^{-2}$ & cm$^{-2}$ & cm$^{-2}$ & cm$^{-2}$ & &  \\
\hline
\endfirsthead
\caption{continued.}\\
\hline\hline
Source & RA (J2000) & Dec (J2000) & $\theta_\mathrm{beam}$ & $T_\mathrm{ex}$ & $N_\mathrm{^{13}CH_3OH}$ & $N_\mathrm{CH_3^{18}OH}$ & $N_\mathrm{CH_3OH}$ & $N_\mathrm{CH_2DOH}$ & $N_\mathrm{CHD_2OH}$ & $\rm (D/H)_{CH_3OH}$\tablefootmark{1} & $\rm (D/H)_{CH_2DOH}$\tablefootmark{2} \\
& & & $^{\prime\prime}$ & K  & cm$^{-2}$ & cm$^{-2}$ & cm$^{-2}$ & cm$^{-2}$ & cm$^{-2}$ & & \\
\hline
\endhead
\hline
\endfoot
86213A & 18:26:48.92 & -12:26:24.51 & 1.23 & 150 & $<$3.3(15) & $<$1.0(15) &  0.4--14.8(16) & $<$2.1(15) & $<$2.4(15) &  -- &  -- \\
86213B & 18:26:47.96 & -12:26:20.73 & 1.23 & 150 & $<$4.0(15) & $<$1.6(15) &  0.3--17.9(16) & $<$1.5(15) & $<$4.0(15) &  -- &  -- \\
86213C & 18:26:48.73 & -12:26:25.98 & 1.23 & 150 & $<$1.9(15) & $<$1.1(15) &  0.5--8.7(16) & $<$1.7(15) & $<$3.0(15) &  -- &  -- \\
81635A & 18:25:00.82 & -13:15:34.46 & 1.22 & 150 & $<$3.8(15) & $<$2.8(15) &  0.1--17.7(16) & $<$2.5(15) & $<$4.7(15) &  -- &  -- \\
81635B & 18:25:01.01 & -13:15:38.57 & 1.22 & 150 & $<$2.7(15) & $<$1.5(15) & $<$1.4(15) & $<$3.1(15) & $<$2.2(15) &  -- &  -- \\
81635C & 18:25:01.65 & -13:15:28.99 & 1.22 & 150 & $<$10.0(15) & $<$1.5(15) & $<$4.0(15) & $<$3.6(15) & $<$4.0(15) &  -- &  -- \\
83968A & 18:25:10.59 & -12:42:22.16 & 1.23 & 150 & $<$2.2(15) & $<$1.4(15) &  0.1--10.9(16) & $<$1.5(15) & $<$5.2(15) &  -- &  -- \\
83968B & 18:25:10.69 & -12:42:26.14 & 1.23 & 150 & $<$3.0(15) & $<$1.2(15) & $<$1.5(15) & $<$1.7(15) & $<$5.1(15) &  -- &  -- \\
83968C & 18:25:10.82 & -12:42:24.68 & 1.23 & 150 & $<$2.3(15) & $<$9.9(14) &  0.1--11.3(16) & $<$1.9(15) & $<$3.2(15) &  -- &  -- \\
83968D & 18:25:10.62 & -12:42:19.43 & 1.23 & 150 & $<$2.3(15) & $<$1.2(15) & $<$1.3(15) & $<$3.0(15) & $<$5.2(15) &  -- &  -- \\
83968E & 18:25:10.65 & -12:42:24.74 & 1.23 & 150 & $<$2.3(15) & $<$1.2(15) & $<$1.3(15) & $<$2.4(15) & $<$5.0(15) &  -- &  -- \\
101899 C1 & 18:34:40.29 & -09:00:38.44 & 1.25 & 150 &  1.5$\pm$0.3(16) & $<$4.0(15) &  6.8$\pm$3.6(17) &  8.0$\pm$4.0(15) &  3.0$\pm$1.5(15) &  3.9$\pm$2.9(-3) &  3.8$\pm$2.7(-1) \\
101899 C2 & 18:34:40.29 & -09:00:38.44 & 1.25 & 150 &  6.7$\pm$1.5(15) & $<$4.0(15) &  2.9$\pm$1.6(17) & $<$6.0(15) & $<$3.0(15) & $<$1.5(-2) &  -- \\
103421 & 18:33:23.98 & -08:33:31.92 & 1.24 & 150 & $<$4.3(15) & $<$1.0(15) &  0.1--1.9(17) & $<$3.3(15) & $<$2.4(15) &  -- &  -- \\
106756A & 18:34:23.98 & -07:54:48.26 & 1.23 & 150 & $<$2.5(15) & $<$2.7(15) &  0.2--11.2(16) & $<$3.5(15) & $<$3.8(15) &  -- &  -- \\
106756B & 18:34:25.55 & -07:54:46.39 & 1.23 & 150 & $<$4.0(15) & $<$2.0(15) & $<$4.2(15) & $<$3.5(15) & $<$6.6(15) &  -- &  -- \\
106756C & 18:34:25.59 & -07:54:43.11 & 1.23 & 150 & $<$1.1(16) & $<$2.6(15) & $<$4.3(15) & $<$3.5(15) & $<$8.0(15) &  -- &  -- \\
126120A & 18:42:37.55 & -04:02:05.17 & 1.17 & 150 & $<$1.9(15) & $<$5.7(14) &  0.5--9.1(16) & $<$1.7(15) & $<$2.4(15) &  -- &  -- \\
126120B & 18:42:37.66 & -04:02:07.27 & 1.17 & 150 & $<$2.0(15) & $<$7.8(14) &  0.3--9.5(16) & $<$1.8(15) & $<$4.1(15) &  -- &  -- \\
126120C & 18:42:36.85 & -04:02:17.66 & 1.17 & 150 & $<$3.8(15) & $<$1.8(15) & $<$5.7(15) & $<$4.2(15) & $<$6.3(15) &  -- &  -- \\
126120D & 18:42:37.14 & -04:02:02.37 & 1.17 & 150 & $<$3.4(15) & $<$9.8(14) & $<$1.0(15) & $<$2.0(15) & $<$2.9(15) &  -- &  -- \\
126348 & 18:42:51.98 & -03:59:54.37 & 1.16 & 150 &  1.1$\pm$0.3(16) &  3.0$\pm$2.0(15) &  9.4$\pm$6.9(17) & $<$5.0(15) & $<$1.7(15) & $<$6.9(-3) &  -- \\
565926A & 08:02:42.97 & -34:31:48.77 & 0.58 & 150 & $<$5.1(15) & $<$2.5(15) & $<$2.5(15) & $<$4.1(15) & $<$6.4(15) &  -- &  -- \\
565926B & 08:02:42.94 & -34:31:49.96 & 0.58 & 150 & $<$5.0(15) & $<$2.4(15) & $<$4.5(15) & $<$6.2(15) & $<$4.9(15) &  -- &  -- \\
565926C & 08:02:42.72 & -34:31:49.61 & 0.58 & 150 & $<$5.1(15) & $<$2.5(15) & $<$2.5(15) & $<$4.1(15) & $<$4.5(15) &  -- &  -- \\
586092A & 08:32:08.70 & -43:13:45.44 & 0.92 & 75 &  5.1$\pm$1.1(15) &  1.5$\pm$0.3(15) &  8.0$\pm$4.5(17) & $<$1.0(16) & $<$3.6(15) & $<$9.4(-3) &  -- \\
586092B & 08:32:08.48 & -43:13:49.28 & 0.92 & 150 & $<$2.9(15) & $<$1.6(15) &  0.1--2.1(17) & $<$5.6(15) & $<$4.3(15) &  -- &  -- \\
586092C & 08:32:09.06 & -43:13:43.28 & 0.92 & 150 & $<$2.8(15) & $<$1.7(15) & $<$2.5(15) & $<$7.6(15) & $<$4.8(15) &  -- &  -- \\
615590 C1 & 09:24:41.96 & -52:02:08.04 & 0.64 & 200 &  2.0$\pm$0.5(16) & $<$5.0(15) &  1.4$\pm$0.7(18) &  3.0$\pm$1.5(16) & $<$1.0(16) &  7.1$\pm$5.2(-3) & $<$6.7(-1) \\
615590 C2 & 09:24:41.96 & -52:02:08.04 & 0.64 & 150 &  1.5$\pm$0.3(16) & $<$5.0(15) &  1.1$\pm$0.6(18) & $<$1.0(16) & $<$1.0(16) & $<$6.7(-3) &  -- \\
640076A & 10:20:15.66 & -58:03:56.32 & 0.87 & 150 & $<$4.1(15) & $<$1.2(15) &  0.1--2.9(17) & $<$4.3(15) & $<$5.3(15) &  -- &  -- \\
640076B & 10:20:15.60 & -58:03:53.47 & 0.87 & 150 & $<$4.1(15) & $<$2.6(15) &  0.1--2.9(17) & $<$3.0(15) & $<$6.7(15) &  -- &  -- \\
644284A & 10:31:29.78 & -58:02:19.27 & 0.86 & 150 &  6.2$\pm$1.4(15) &  1.5$\pm$0.3(15) &  4.2$\pm$2.2(17) & $<$4.0(15) & $<$3.8(15) & $<$6.7(-3) &  -- \\
644284B & 10:31:29.63 & -58:02:18.82 & 0.86 & 100 &  8.2$\pm$1.8(15) &  1.5$\pm$0.3(15) &  7.8$\pm$4.3(17) & $<$3.0(15) & $<$3.9(15) & $<$2.9(-3) &  -- \\
693050 & 12:35:35.05 & -63:02:31.19 & 0.99 & 200 &  1.4$\pm$0.3(16) &  3.1$\pm$0.7(15) &  1.3$\pm$0.8(18) &  1.1$\pm$0.3(16) & $<$3.0(15) &  2.7$\pm$1.7(-3) & $<$3.9(-1) \\
695243 & 12:43:31.51 & -62:36:13.25 & 0.98 & 150 & $<$4.0(15) & $<$1.2(15) & $<$1.8(15) & $<$5.9(15) & $<$7.9(15) &  -- &  -- \\
704792 & 13:11:14.14 & -62:45:06.80 & 1.29 & 150 & $<$4.5(15) & $<$8.2(14) & $<$1.2(15) & $<$4.5(15) & $<$2.2(15) &  -- &  -- \\
705768 & 13:12:36.17 & -62:33:34.43 & 0.87 & 150 &  9.3$\pm$2.2(15) &  3.0$\pm$1.5(15) &  1.3$\pm$0.8(18) & $<$6.3(15) & $<$4.9(15) & $<$3.7(-3) &  -- \\
706733A & 13:14:22.78 & -62:45:59.48 & 0.87 & 150 & $<$2.7(15) & $<$1.8(15) &  0.2--16.4(16) & $<$4.8(15) & $<$2.9(15) &  -- &  -- \\
706733B & 13:14:22.99 & -62:45:54.35 & 0.87 & 150 & $<$6.7(15) & $<$1.5(15) & $<$2.0(15) & $<$6.6(15) & $<$2.9(15) &  -- &  -- \\
706733C & 13:14:23.07 & -62:45:47.54 & 0.87 & 150 & $<$5.6(15) & $<$1.8(15) & $<$2.5(15) & $<$4.9(15) & $<$9.0(15) &  -- &  -- \\
706785A & 13:14:26.92 & -62:44:29.72 & 0.88 & 150 & $<$3.2(15) & $<$2.1(15) &  0.1--1.9(17) & $<$6.2(15) & $<$2.6(15) &  -- &  -- \\
706785B & 13:14:26.55 & -62:44:31.80 & 0.88 & 150 & $<$3.3(15) & $<$2.9(15) &  0.0--2.0(17) & $<$6.0(15) & $<$3.6(15) &  -- &  -- \\
706785C & 13:14:26.38 & -62:44:30.24 & 0.88 & 150 & $<$2.7(15) & $<$1.3(15) & $<$1.7(15) & $<$5.9(15) & $<$9.0(15) &  -- &  -- \\
706785D & 13:14:25.64 & -62:44:30.36 & 0.88 & 150 & $<$4.4(15) & $<$3.2(15) &  0.1--2.7(17) & $<$6.1(15) & $<$3.7(15) &  -- &  -- \\
707948 & 13:16:43.19 & -62:58:32.83 & 0.88 & 150 &  1.8$\pm$0.4(17) &  1.5$\pm$0.3(16) &  1.1$\pm$0.6(19) &  2.0$\pm$1.0(17) &  6.5$\pm$4.8(16) &  6.0$\pm$4.3(-3) &  3.2$\pm$2.9(-1) \\
717461A & 13:43:01.68 & -62:08:51.42 & 1.29 & 150 &  7.6$\pm$1.9(15) &  2.6$\pm$0.8(15) &  1.1$\pm$0.4(18) & $<$3.7(15) & $<$2.5(15) & $<$2.0(-3) &  -- \\
717461B & 13:43:01.74 & -62:08:55.34 & 1.29 & 150 & $<$2.2(15) & $<$1.9(15) & $<$2.8(15) & $<$3.8(15) & $<$3.6(15) &  -- &  -- \\
721992 & 13:51:58.27 & -61:15:41.04 & 0.85 & 150 & $<$4.4(15) & $<$1.9(15) &  0.5--2.5(17) & $<$7.3(15) & $<$5.5(15) &  -- &  -- \\
724566 & 13:59:30.92 & -61:48:38.27 & 0.83 & 150 &  2.9$\pm$0.6(16) &  1.0$\pm$0.2(16) &  4.0$\pm$2.3(18) &  2.0$\pm$1.0(16) & $<$9.6(15) &  1.7$\pm$1.3(-3) & $<$9.6(-1) \\
732038 & 14:13:15.05 & -61:16:53.19 & 0.82 & 150 & $<$7.6(15) & $<$4.8(15) &  1.2--4.2(17) & $<$6.8(15) & $<$7.5(15) &  -- &  -- \\
744757A & 14:45:26.35 & -59:49:15.55 & 1.30 & 150 &  1.4$\pm$0.4(16) &  4.1$\pm$0.9(15) &  1.7$\pm$1.0(18) &  8.0$\pm$4.0(15) & $<$2.5(15) &  1.6$\pm$1.2(-3) & $<$6.3(-1) \\
744757B & 14:45:26.16 & -59:49:19.87 & 1.30 & 150 & $<$1.9(15) & $<$4.1(15) &  1.3--11.3(16) & $<$2.9(15) & $<$2.3(15) &  -- &  -- \\
759150A & 15:10:43.13 & -57:44:49.63 & 1.29 & 150 & $<$1.9(15) & $<$1.2(15) &  0.3--10.1(16) & $<$5.0(15) & $<$2.3(15) &  -- &  -- \\
759150B & 15:10:43.52 & -57:44:44.82 & 1.29 & 150 & $<$2.4(15) & $<$9.6(14) & $<$1.5(15) & $<$5.0(15) & $<$2.8(15) &  -- &  -- \\
759150C & 15:10:44.48 & -57:44:47.33 & 1.29 & 150 & $<$2.2(15) & $<$9.6(14) & $<$8.6(14) & $<$5.0(15) & $<$2.3(15) &  -- &  -- \\
759150D & 15:10:42.71 & -57:44:52.85 & 1.29 & 150 & $<$5.0(15) & $<$9.7(14) & $<$1.2(15) & $<$4.9(15) & $<$2.3(15) &  -- &  -- \\
759150E & 15:10:44.10 & -57:44:52.03 & 1.29 & 150 & $<$2.1(15) & $<$9.6(14) & $<$9.9(14) & $<$4.8(15) & $<$2.3(15) &  -- &  -- \\
767784 & 15:29:19.31 & -56:31:22.02 & 1.29 & 100 &  3.9$\pm$0.5(16) &  7.2$\pm$1.6(15) &  2.5$\pm$1.4(18) &  1.0$\pm$0.4(16) & $<$3.3(15) &  1.4$\pm$1.0(-3) & $<$5.3(-1) \\
800287 & 16:11:26.57 & -51:41:57.14 & 0.80 & 100 &  1.5$\pm$0.3(16) &  6.2$\pm$1.4(15) &  1.8$\pm$1.1(18) &  1.0$\pm$0.5(16) & $<$1.3(16) &  1.9$\pm$1.5(-3) & $<$2.7(0) \\
854214A & 16:52:32.74 & -43:23:49.60 & 1.26 & 150 & $<$2.2(15) & $<$1.7(15) &  0.7--9.5(16) & $<$3.8(15) & $<$5.0(15) &  -- &  -- \\
854214B & 16:52:33.02 & -43:23:50.26 & 1.26 & 150 & $<$5.0(15) & $<$7.8(14) &  0.1--2.1(17) & $<$3.8(15) & $<$2.7(15) &  -- &  -- \\
863312A & 17:02:08.36 & -41:46:56.89 & 0.83 & 150 & $<$8.9(15) & $<$3.7(15) & $<$3.8(15) & $<$2.0(16) & $<$1.4(16) &  -- &  -- \\
863312B & 17:02:09.14 & -41:46:45.04 & 0.83 & 150 & $<$3.7(15) & $<$1.9(15) & $<$2.0(15) & $<$6.0(15) & $<$4.2(15) &  -- &  -- \\
865468A C1 & 17:05:10.90 & -41:29:06.99 & 1.23 & 100 &  1.3$\pm$0.3(17) &  3.1$\pm$0.7(16) &  1.0$\pm$0.6(19) &  1.5$\pm$0.8(17) &  4.0$\pm$2.0(16) &  4.9$\pm$3.8(-3) &  2.7$\pm$1.9(-1) \\
865468A C2 & 17:05:10.90 & -41:29:06.99 & 1.23 & 150 &  5.1$\pm$1.1(16) & $<$3.0(15) &  2.5$\pm$1.4(18) &  6.0$\pm$3.0(16) & $<$2.0(16) &  7.9$\pm$5.7(-3) & $<$6.7(-1) \\
865468B & 17:05:11.22 & -41:29:07.65 & 1.24 & 150 &  1.3$\pm$0.2(16) &  2.6$\pm$0.6(15) &  8.6$\pm$5.0(17) &  6.5$\pm$2.0(15) & $<$5.0(15) &  2.5$\pm$1.7(-3) & $<$1.1(0) \\
865468C & 17:05:11.12 & -41:29:03.47 & 1.24 & 150 &  5.0$\pm$1.9(15) &  1.0$\pm$0.2(15) &  3.4$\pm$2.0(17) & $<$7.5(15) & $<$5.8(15) & $<$1.8(-2) &  -- \\
876288 & 17:11:51.02 & -39:09:29.18 & 0.81 & 150 &  8.9$\pm$2.9(15) &  3.1$\pm$0.7(15) &  5.7$\pm$3.9(17) & $<$6.0(15) & $<$5.0(15) & $<$1.2(-2) &  -- \\
881427A & 17:20:06.31 & -38:57:15.18 & 1.23 & 150 &  9.3$\pm$1.8(16) &  2.6$\pm$0.7(16) &  1.1$\pm$0.4(19) &  8.3$\pm$1.6(16) &  1.9$\pm$0.7(16) &  2.5$\pm$1.1(-3) &  2.3$\pm$1.0(-1) \\
881427B & 17:20:06.46 & -38:57:11.44 & 1.23 & 300 &  1.0$\pm$0.2(16) &  8.2$\pm$1.8(15) &  3.4$\pm$1.9(18) &  1.5$\pm$0.8(16) & $<$1.0(16) &  1.5$\pm$1.1(-3) & $<$1.3(0) \\
881427C & 17:20:06.12 & -38:57:15.84 & 1.23 & 150 &  7.4$\pm$1.6(16) &  2.6$\pm$0.5(16) &  1.1$\pm$0.4(19) &  5.5$\pm$1.2(16) &  1.2$\pm$0.5(16) &  1.7$\pm$0.7(-3) &  2.2$\pm$0.9(-1) \\
G023.3891+00.1851 & 18:33:14.32 & -08:23:57.82 & 1.24 & 200 &  7.9$\pm$2.2(15) &  2.5$\pm$1.2(15) &  7.9$\pm$4.5(17) &  9.5$\pm$3.0(15) & $<$5.0(15) &  4.0$\pm$2.6(-3) & $<$7.6(-1) \\
G023.6566-00.1273 & 18:34:51.57 & -08:18:21.81 & 1.24 & 150 &  1.4$\pm$0.4(16) & $<$5.0(15) &  7.1$\pm$2.5(17) &  2.5$\pm$1.2(16) & $<$5.0(15) &  1.2$\pm$0.7(-2) & $<$4.0(-1) \\
G025.6498+01.0491 & 18:34:20.92 & -05:59:42.08 & 1.17 & 150 &  3.0$\pm$0.6(16) &  8.5$\pm$2.7(15) &  3.4$\pm$1.4(18) &  1.5$\pm$0.8(16) &  5.0$\pm$2.5(15) &  1.5$\pm$1.0(-3) &  3.3$\pm$2.4(-1) \\
G030.1981-00.1691 & 18:47:03.05 & -02:30:36.30 & 0.60 & 150 &  5.1$\pm$1.1(15) & $<$1.8(15) &  2.2$\pm$1.2(17) & $<$1.5(16) & $<$1.6(16) & $<$4.7(-2) &  -- \\
G233.8306-00.1803 & 07:30:16.73 & -18:35:49.06 & 0.81 & 150 & $<$3.7(15) & $<$3.0(15) & $<$2.6(15) & $<$6.5(15) & $<$4.4(15) &  -- &  -- \\
G305.2017+00.2072A1 & 13:11:10.45 & -62:34:38.60 & 1.30 & 150 &  1.2$\pm$0.3(16) &  2.0$\pm$0.5(15) &  8.5$\pm$4.8(17) &  5.0$\pm$2.5(15) & $<$2.3(15) &  2.0$\pm$1.5(-3) & $<$9.1(-1) \\
G305.2017+00.2072A2 & 13:11:13.12 & -62:34:42.74 & 1.30 & 150 & $<$8.8(15) & $<$5.0(15) &  0.1--5.3(17) & $<$8.0(15) & $<$1.2(16) &  -- &  -- \\
G310.0135+00.3892 & 13:51:37.88 & -61:39:07.74 & 1.30 & 150 & $<$2.3(15) & $<$2.6(15) &  0.3--13.5(16) & $<$7.9(15) & $<$2.9(15) &  -- &  -- \\
G314.3197+00.1125 & 14:26:26.25 & -60:38:31.26 & 1.30 & 150 & $<$2.4(15) & $<$2.2(15) &  0.8--1.4(17) & $<$1.0(16) & $<$6.0(15) &  -- &  -- \\
G316.6412-00.0867 & 14:44:18.35 & -59:55:11.28 & 1.29 & 100\tablefootmark{3} &  2.8$\pm$0.4(16) &  8.6$\pm$2.7(15) &  3.5$\pm$1.5(18) &  2.8$\pm$0.5(16) &  4.8$\pm$3.0(15) &  2.7$\pm$1.2(-3) &  1.7$\pm$1.1(-1) \\
G318.0489+00.0854B & 14:53:42.64 & -59:08:53.02 & 1.30 & 150 &  1.7$\pm$0.3(16) &  4.9$\pm$1.9(15) &  1.9$\pm$0.9(18) &  8.0$\pm$4.0(15) & $<$3.0(15) &  1.4$\pm$1.0(-3) & $<$7.5(-1) \\
G318.9480-00.1969A1 & 15:00:55.28 & -58:58:52.60 & 1.29 & 100\tablefootmark{3} &  8.5$\pm$1.2(16) &  2.2$\pm$0.4(16) &  9.6$\pm$3.2(18) &  7.0$\pm$1.1(16) &  1.4$\pm$0.7(16) &  2.4$\pm$0.9(-3) &  2.0$\pm$1.1(-1) \\
G318.9480-00.1969A2 & 15:00:55.23 & -58:58:55.88 & 1.29 & 150 & $<$2.6(15) & $<$7.8(14) &  0.4--16.0(16) & $<$6.5(15) & $<$2.3(15) &  -- &  -- \\
G323.7399-00.2617B1 & 15:31:45.64 & -56:30:50.16 & 1.28 & 150 &  8.7$\pm$2.0(15) &  2.0$\pm$0.5(15) &  7.5$\pm$4.3(17) & $<$2.8(15) & $<$3.0(15) & $<$2.9(-3) &  -- \\
G323.7399-00.2617B2 & 15:31:45.45 & -56:30:49.84 & 1.28 & 125 &  8.2$\pm$1.8(16) &  1.5$\pm$0.3(16) &  5.7$\pm$3.3(18) &  2.0$\pm$1.0(16) &  4.0$\pm$2.0(15) &  1.2$\pm$0.9(-3) &  2.0$\pm$1.4(-1) \\
G323.7399-00.2617B3 & 15:31:45.73 & -56:30:51.93 & 1.28 & 150 & $<$2.5(15) & $<$6.0(14) &  1.4--13.4(16) & $<$4.7(15) & $<$2.3(15) &  -- &  -- \\
G323.7399-00.2617B4 & 15:31:45.94 & -56:30:51.34 & 1.28 & 150 & $<$2.6(15) & $<$6.9(14) &  1.2--14.2(16) & $<$2.7(15) & $<$2.8(15) &  -- &  -- \\
G323.7399-00.2617B5 & 15:31:45.62 & -56:30:45.62 & 1.28 & 150 & $<$2.6(15) & $<$5.8(14) &  0.1--13.9(16) & $<$5.0(15) & $<$2.7(15) &  -- &  -- \\
G323.7399-00.2617B6 & 15:31:45.84 & -56:30:47.68 & 1.28 & 150 & $<$3.9(15) & $<$1.2(15) &  0.1--2.1(17) & $<$2.6(15) & $<$3.7(15) &  -- &  -- \\
G323.7399-00.2617B7 & 15:31:45.91 & -56:30:46.10 & 1.28 & 150 & $<$1.7(15) & $<$1.2(15) &  0.1--9.3(16) & $<$3.9(15) & $<$1.6(15) &  -- &  -- \\
G327.1192+00.5103 & 15:47:32.72 & -53:52:38.60 & 0.81 & 100 &  4.0$\pm$0.9(16) &  9.2$\pm$2.0(15) &  2.9$\pm$1.7(18) &  3.5$\pm$1.8(16) & $<$5.0(15) &  4.0$\pm$3.1(-3) & $<$2.9(-1) \\
G343.1261-00.0623 & 16:58:17.22 & -42:52:07.54 & 1.25 & 100 &  8.8$\pm$2.7(15) & $<$7.0(15) &  5.0$\pm$1.8(17) & $<$4.0(16) &  4.0$\pm$2.0(15) & $<$4.1(-2) & $>$5.0(-2) \\
G345.5043+00.3480 C1 & 17:04:22.89 & -40:44:23.06 & 1.25 & 125 &  5.1$\pm$1.1(16) &  1.0$\pm$0.2(16) &  4.0$\pm$2.3(18) &  5.0$\pm$2.5(16) &  1.0$\pm$0.5(16) &  4.2$\pm$3.2(-3) &  2.0$\pm$1.4(-1) \\
G345.5043+00.3480 C2 & 17:04:22.89 & -40:44:23.06 & 1.25 & 150 &  3.1$\pm$0.7(16) & $<$1.0(16) &  1.7$\pm$0.9(18) &  3.0$\pm$1.5(16) & $<$1.0(16) &  5.9$\pm$4.3(-3) & $<$6.7(-1) \\
G348.7342-01.0359B1 & 17:20:07.08 & -38:57:11.22 & 1.23 & 150 & $<$3.4(15) & $<$9.9(14) &  1.3--17.7(16) & $<$1.2(16) & $<$3.9(15) &  -- &  -- \\
G348.7342-01.0359B2 & 17:20:07.26 & -38:57:09.82 & 1.23 & 150 & $<$2.3(15) & $<$1.7(15) &  0.1--11.9(16) & $<$3.6(15) & $<$4.1(15) &  -- &  -- \\
G348.7342-01.0359B3 & 17:20:07.38 & -38:57:10.15 & 1.23 & 150 & $<$4.0(15) & $<$1.5(15) & $<$1.4(15) & $<$1.1(16) & $<$3.8(15) &  -- &  -- \\
\end{longtable}
\tablefoot{
The coordinates mark the position from which the spectra were extracted. All column densities are derived for the reported $T_\mathrm{ex}$ and assuming the source size is equal to the size of the beam (i.e., beam dilution = 1). The column density of CH$_3$OH is derived from the CH$_3^{18}$OH and $^{13}$CH$_3$OH isotopologues when these are detected. If neither isotopologue is detected but CH$_3$OH is, a range in column densities is presented where the lower limit it the column density derived from CH$_3$OH and the upper limit is scaled from the upper limit on $^{13}$CH$_3$OH. If CH$_3$OH itself is also not detected, an upper limit is directly derived from its spectrum.
\\
\tablefoottext{1}{Derived from $N_\mathrm{CH_2DOH}$ using Eq.~\eqref{eq:CH3OH_D_H_CH2DOH}.}
\tablefoottext{2}{Derived from $N_\mathrm{CHD_2OH}$ using Eq.~\eqref{eq:CH2DOH_D_H_CHD2OH}.}
\tablefoottext{3}{$T_\mathrm{ex}$ is set to 150~K for deuterated isotopologues and to 100~K for the other isotopologues.}

}
}
\end{landscape}
\renewcommand*{\arraystretch}{1.0}

\section{Methanol D/H ratios of sources in the literature}
\label{app:D_H_literature}
\begin{longtable}[c]{lllll}
\caption{The $\rm (D/H)_{CH_3OH}$ and $\rm (D/H)_{CH_2DOH}$ ratios taken from the literature that are included in Figs.~\ref{fig:CH2DOH_CH3OH_D_H_names} and \ref{fig:CHD2OH_CH2DOH_D_H_names} .\label{tab:D_H_literature}}\\
\hline\hline
Source & Type\tablefootmark{(1)} & $\rm (D/H)_{CH_3OH}$ & $\rm (D/H)_{CH_2DOH}$ & Refs \\
& & & &  \\
\hline
\endfirsthead
\caption{continued.}\\
\hline\hline
Source & Type\tablefootmark{(1)} & $\rm (D/H)_{CH_3OH}$ & $\rm (D/H)_{CH_2DOH}$ & Refs \\
& & & &  \\
\hline
\endhead
\hline
\endfoot
B1-c & LMP & 2.8$\pm$0.9(-2) &  1.3$\pm$0.2(-1) & 1,2 \\
Serpens S68N & LMP & 1.4$\pm$0.6(-2) &  1.2$\pm$0.5(-1) & 1,2  \\
B1-bS & LMP & $<$1.8(-2) &  -- & 1,2  \\
HH212 & LMP & 3.8$\pm$2.3(-2) &  -- & 3 \\
	  &     & 8.1$\pm$3.0(-3) &  -- & 4 \\
IRAS 16293A & LMP & 2.8$\pm$1.2(-2) &  2.0$\pm$0.7(-1) & 5,6 \\
IRAS 16293B & LMP & 2.4$\pm$0.9(-2) &  2.5$\pm$0.9(-1) & 6,7 \\
IRAS 2A & LMP & 1.9$\pm$1.0(-2) &  7.0$\pm$2.6(-1) & 8 \\
IRAS 4A & LMP & 1.4$\pm$0.8(-2) &  5.6$\pm$2.2(-1) & 8 \\
L483 & LMP & 7.8$\pm$3.3(-3) &  -- & 9 \\
BHR71 & LMP & 9.6$\pm$4.1(-3) &  -- & 10 \\
Ser-emb 1 & LMP & $<$2.0(-1) &  -- & 11 \\
Ser-emb 11W & LMP & $<$4.3(-2) &  -- & 12 \\
HOPS-108 & LMP & 7.0$\pm$3.8(-3) &  -- & 13 \\
G192.12–11.10 & LMP & 9.8$\pm$5.8(-3) &  -- & 14 \\
G205.46–14.56S1–A & LMP & 6.7$\pm$3.3(-3) &  -- & 14 \\
G208.68–19.20N1 & LMP & 1.3$\pm$0.6(-2) &  -- & 14 \\
G210.49–19.79W–A & LMP & 1.4$\pm$0.6(-2) &  -- & 14 \\
G211.47–19.27S & LMP & 1.7$\pm$0.4(-2) &  -- & 14 \\
V883 Ori & LMP & 4.8$\pm$1.0(-2) &  -- & 15 \\
Serpens SMM1-a & IMP & 3.6$\pm$2.1(-3) &  -- & 16 \\
NGC 7192 FIRS2 & IMP & 1.9$\pm$0.8(-3) &  -- & 17 \\
Cep E-A & IMP & 1.2$\pm$0.4(-2) &  -- & 18 \\
Sgr B2(N2) & HMP & 4.0$\pm$1.7(-4) &  -- & 19,20 \\
NGC6334I MM1 I & HMP & 3.0$\pm$1.7(-4) &  -- & 21 \\
NGC6334I MM1 II & HMP & 1.8$\pm$0.9(-4) &  -- & 21 \\
NGC6334I MM1 III & HMP & 3.2$\pm$1.8(-4) &  -- & 21 \\
NGC6334I MM1 IV & HMP & 1.6$\pm$1.0(-4) &  -- & 21 \\
NGC6334I MM1 V & HMP & 2.3$\pm$1.1(-4) &  -- & 21 \\
NGC6334I MM2 I & HMP & 6.0$\pm$3.6(-4) &  -- & 21 \\
NGC6334I MM2 II & HMP & 1.5$\pm$0.7(-4) &  -- & 21 \\
NGC6334I MM3 I & HMP & 2.4$\pm$1.1(-4) &  -- & 21 \\
NGC6334I MM3 II & HMP & 2.6$\pm$1.1(-4) &  -- & 21 \\
Orion KL ridge & HMP & 1.7$\pm$0.7(-3) &  -- & 22 \\
Orion KL HC & HMP & $<$1.7(-3) &  -- & 22 \\
CygX-N30 & HMP & $<$1.7(-3) &  -- & 23 \\
L1495-B10 6 & LMPC & 1.4$\pm$0.4(-2) &  -- & 24 \\
L1495-B10 7 & LMPC & 4.8$\pm$1.3(-2) &  -- & 24 \\
L1495-B10 8 & LMPC & 5.2$\pm$1.5(-2) &  -- & 24 \\
L1495-B10 9 & LMPC & 4.6$\pm$1.1(-2) &  -- & 24 \\
L1495-B10 10 & LMPC & 4.2$\pm$1.3(-2) &  -- & 24 \\
L1495-B10 11 & LMPC & $<$1.9(-2) &  -- & 24 \\
L1495-B10 12 & LMPC & 2.0$\pm$0.4(-2) &  -- & 24 \\
L1495-B10 13 & LMPC & $<$3.5(-2) &  -- & 24 \\
L1495-B10 14 & LMPC & $<$1.4(-2) &  -- & 24 \\
L1495-B10 15 & LMPC & 4.0$\pm$1.1(-2) &  -- & 24 \\
L1495-B10 16 & LMPC & 3.0$\pm$1.0(-2) &  -- & 24 \\
L1495-B10 17 & LMPC & 8.8$\pm$2.7(-2) &  -- & 24 \\
L183 & LMPC & 1.3$\pm$0.2(-2) &  -- & 25 \\
L1544 & LMPC & 3.0$\pm$1.3(-2) &  -- & 26 \\
I00117-MM2 & HMSC & $<$2.4(-3) &  -- & 27 \\
AFGL 5142-EC & HMSC & 9.0$\pm$3.8(-4) &  -- & 27 \\
05458-mm3 & HMSC & 2.3$\pm$1.0(-3) &  -- & 27 \\
G034-G2(MM2) & HMSC & 1.4$\pm$0.6(-2) &  -- & 27 \\
G034-F2(MM7) & HMSC & $<$2.5(-3) &  -- & 27 \\
G034-F1(MM8) & HMSC & $<$1.5(-3) &  -- & 27 \\
G034-C1(MM9) & HMSC & $<$8.7(-4) &  -- & 27 \\
I20293-WC & HMSC & $<$1.9(-3) &  -- & 27 \\
I22134-G & HMSC & $<$1.7(-3) &  -- & 27 \\
I22134-B & HMSC & $<$6.7(-3) &  -- & 27 \\
\end{longtable}
\tablefoot{
The $\rm (D/H)_{CH_3OH}$ and $\rm (D/H)_{CH_2DOH}$ ratios are either directly taken from the reported literature or computed using  Eqs.~\eqref{eq:CH3OH_D_H_CH2DOH} and \eqref{eq:CH2DOH_D_H_CHD2OH} using the column densities from the reported literature. A 30\% uncertainty was assumed in cases where no uncertainty was reported. \\
\tablefoottext{1}{LMP: low-mass protostar, IMS: intermediate-mass protostar, HMS: high-mass protostar, LMPC: low-mass prestellar core, HMSC: high-mass starless core.} \\ \\
{\bf References. } 
1: \citet{vanGelder2020};
2: Appendix~\ref{app:CHD2OH_lowmass};
3: \citet{Lee2019_HH212};
4: \citet{Bianchi2017_HH212};
5: \citet{Manigand2020};
6: \citet{Drozdovskaya2022};
7: \citet{Jorgensen2018};
8: \citet{Taquet2019};
9: \citet{Jacobsen2019};
10: \citet{Yang2020};
11: \citet{Martin-domenech2019};
12: \citet{Martin-domenech2021};
13: \citet{Chahine2022};
14: \citet{Hsu2022}; 
15: \citet{Lee2019_V883};
16: \citet{Ligterink2021};
17: \citet{Fuente2014};
18: \citet{Ospina-Zamudio2018};
19: \citet{Belloche2016};
20: \citet{Muller2016};
21: \citet{Bogelund2018};
22: \citet{Neill2013};
23: \citet{vanderWalt2021};
24: \citet{Ambrose2021};
25: \citet{Lattanzi2020};
26: \citet{Bizzocchi2014};
27: \citet{Fontani2015}.
}


\section{Doubly deuterated methanol in B1-c, Serpens S68N, and B1-bS}
\label{app:CHD2OH_lowmass}
Using the database entry of CHD$_2$OH provided by \citet{Drozdovskaya2022}, transitions from CHD$_2$OH can also be searched for in a few COM-rich low-mass protostars. Here, this is done for B1-c, Serpens~S68N (hereafter S68N), and B1-bS from the 2017.1.01174.S ALMA program. The content of oxygen-bearing COMs for these sources was presented by \citet{vanGelder2020}, but no public database entry was yet available for CHD$_2$OH at that time.

\begin{figure*}[h]
\includegraphics[height=4.78cm]{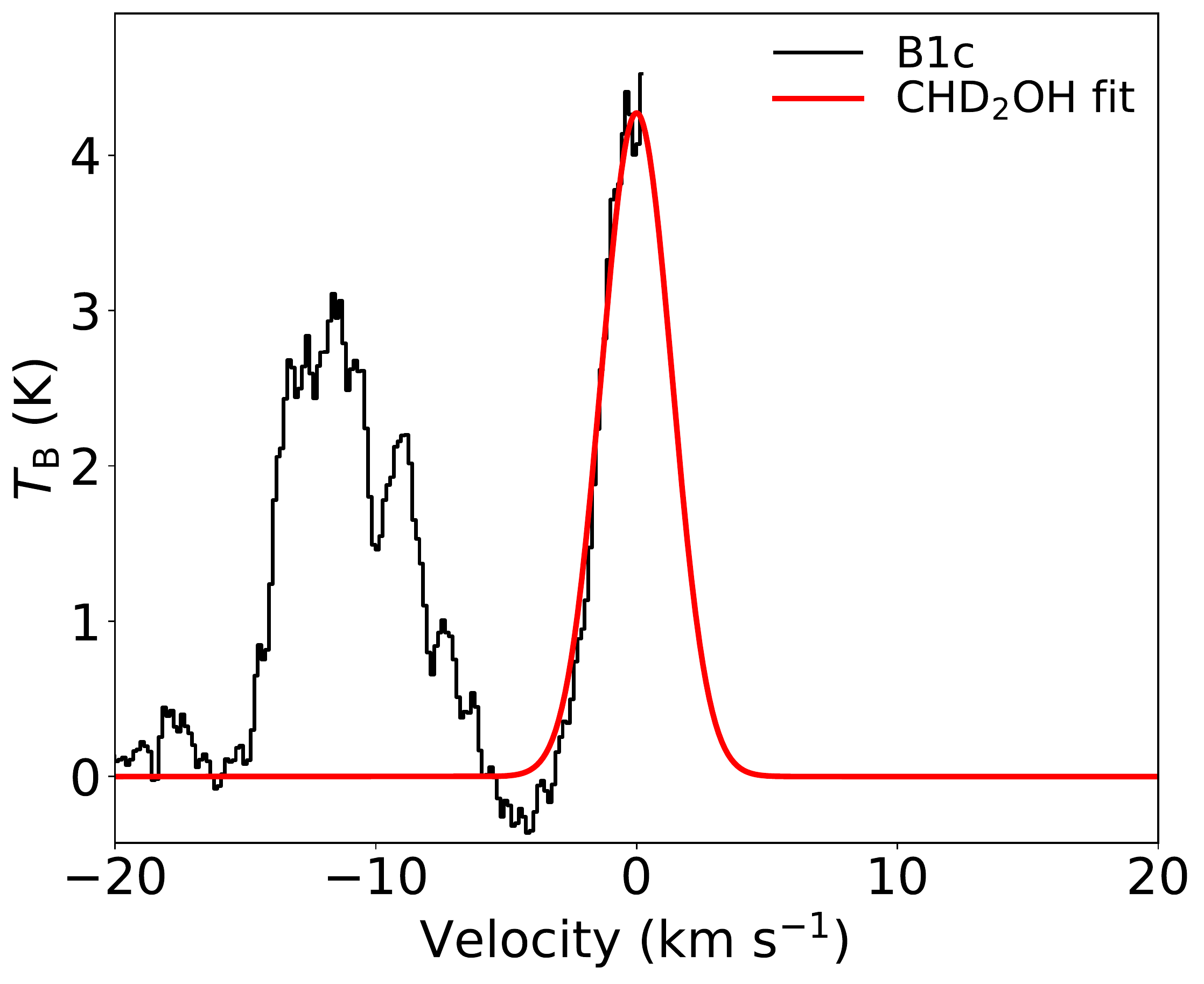}
\includegraphics[height=4.78cm]{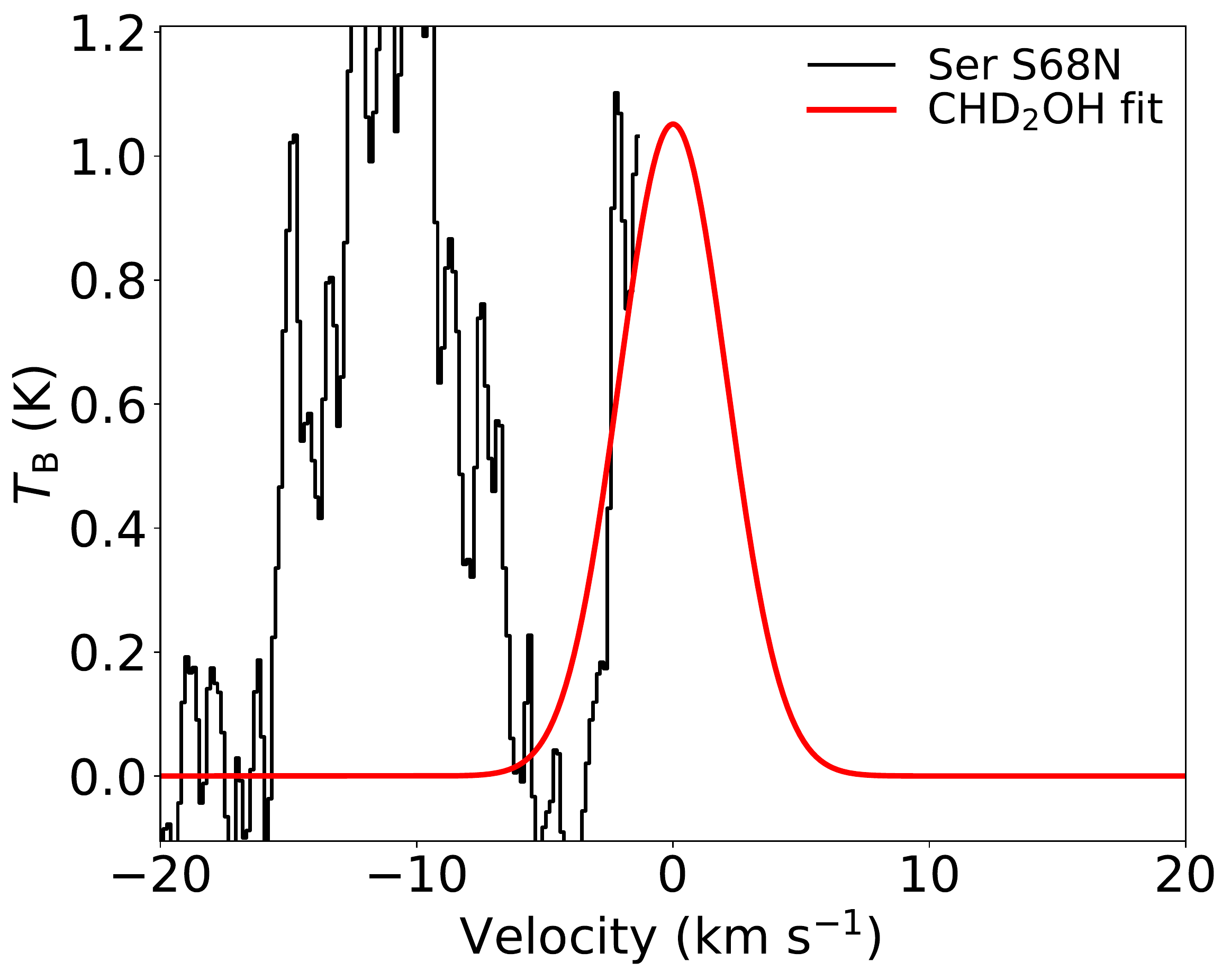}
\includegraphics[height=4.78cm]{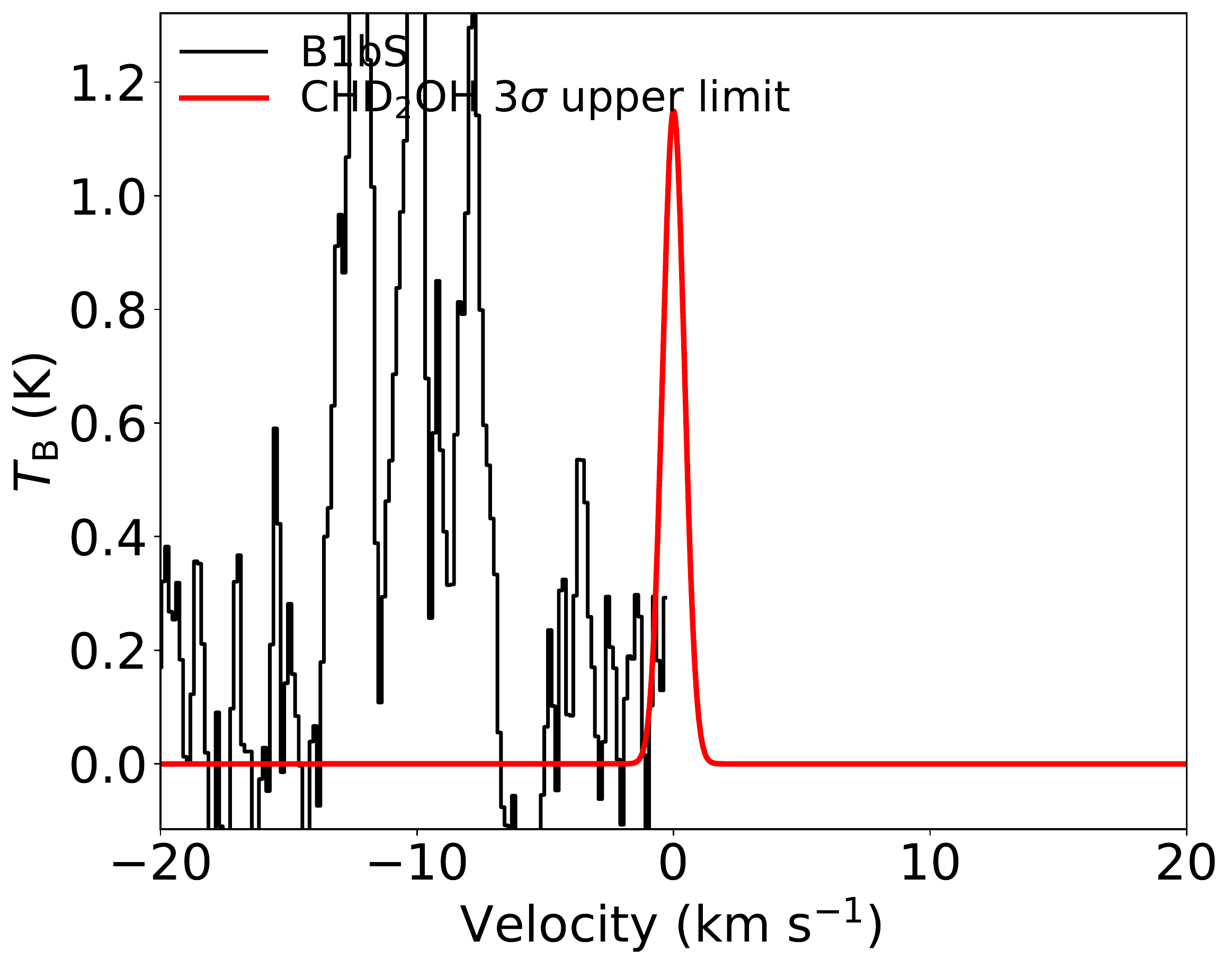}
\caption{
Spectral line fits of CHD$_2$OH $7_{1,2}\,o_1-7_{0,1}\,e_0$ ($E_\mathrm{up} = 68$~K) line for B1-c (left), Serpens~S68N (middle), and B1-bS (right). The data corrected for the $V_\mathrm{lsr}$ are shown in black and the best fit for $T_\mathrm{ex}=150$~K is shown in red. 
}
\label{fig:linefits_B1c_S68N_B1bS}
\end{figure*}

Only one strong transition of CHD$_2$OH ($7_{1,2}\,o_1-7_{0,1}\,e_0$, $E_\mathrm{up} = 68$~K) is available for these sources which lies on the very edge of the covered frequency range. For B1-c and S68N, this transition is detected at the 3$\sigma$ level, but given that only half the line is observed this detection is still tentative. Using the same method for deriving the column densities as used by \citet{vanGelder2020} and assuming an excitation temperature of 150~K, we derive column densities of $2.0\pm0.6\times10^{16}$~cm$^{-2}$ for B1-c, $7.2\pm2.7\times10^{15}$~cm$^{-2}$ for S68N, and $<1.7\times10^{15}$~cm$^{-2}$ for B1-bS. The FWHM was fixed to the average FWHM of those sources of 3.2\kms, 5.5\kms, and 1.0\kms, respectively \citep{vanGelder2020}. The resulting fits are shown in Fig.~\ref{fig:linefits_B1c_S68N_B1bS}. Using these derived column densities and those reported for CH$_3$OH by \citet{vanGelder2020}, the $\rm (D/H)_{CH_2DOH}$ ratios for B1-c and S68N (for B1-bS, both CH$_2$DOH and CHD$_2$OH are not detected) are shown in Fig.~\ref{fig:CHD2OH_CH2DOH_D_H_names} and agree very well with those derived for other low-mass sources as well as with the high-mass sources. 


\newpage
\section{Additional figures}
\begin{figure*}[!h]
\includegraphics[width=\linewidth]{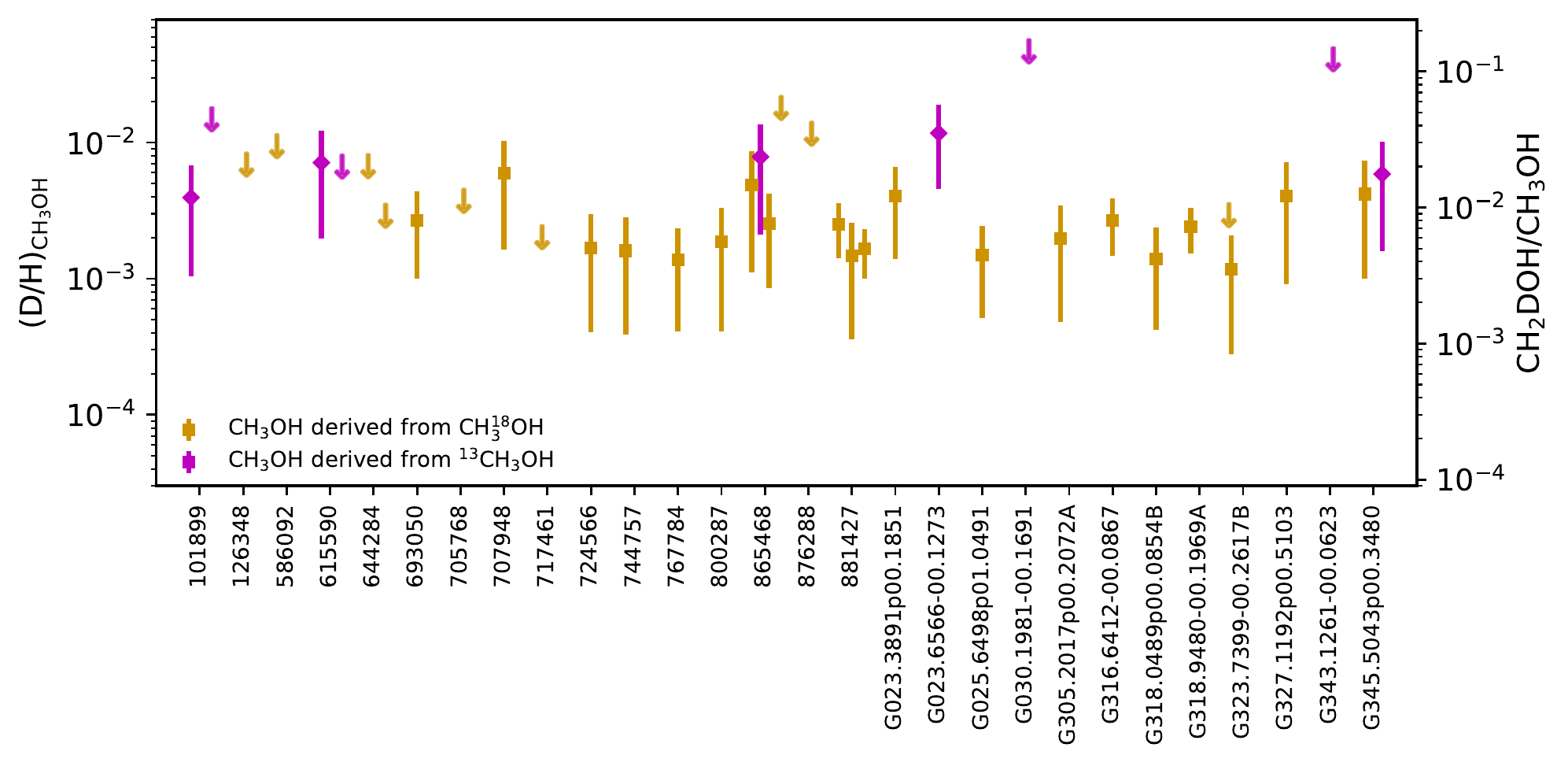}
\caption{
The $\rm (D/H)_{CH_3OH}$ ratios derived from the $N_\mathrm{CH_2DOH}/N_\mathrm{CH_3OH}$ ratios for the ALMAGAL sources presented in this work, indicating whether $N_\mathrm{CH_3OH}$ was derived from the $^{13}$C isotopologue (magenta diamonds) or from the $^{18}$O isotopologue (orange squares). Upper limits are presented as arrows.
}
\label{fig:CH2DOH_CH3OH_D_H_ALMAGAL_isotopologues}
\end{figure*}

\begin{figure}[!h]
\includegraphics[width=\linewidth]{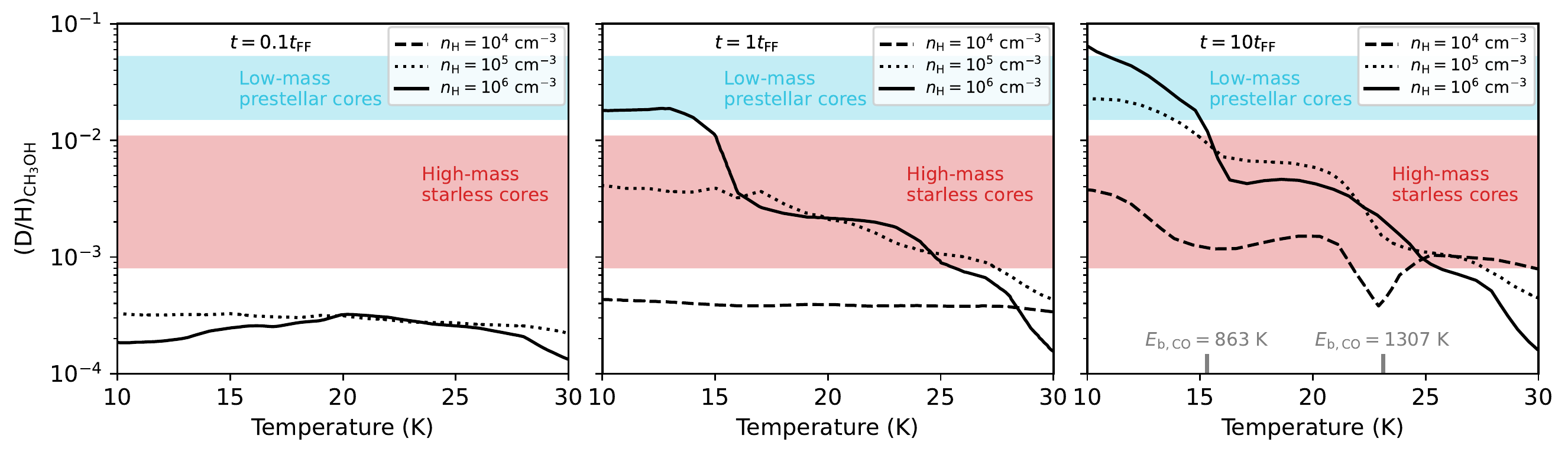}
\caption{Same as Fig.~\ref{fig:D_H_models_HMP,LMP} but now showing the average measured $\rm (D/H)_{CH_3OH}$ for low-mass prestellar cores (light blue) and high-mass starless cores (red). }
\label{fig:D_H_models_HMSC,LMPC}
\end{figure}

\end{appendix}

\end{document}